\documentclass[twocolumn,times]{aastex631}

\usepackage{tabularx}
\usepackage{longtable}
\usepackage{extarrows}
\usepackage{supertabular}
\usepackage{threeparttablex}
\usepackage{graphicx}
\usepackage{dcolumn}
\usepackage{bm}
\usepackage{amsmath}
\usepackage{overpic}
\usepackage{booktabs}%
\usepackage{makecell}
\usepackage{enumerate}
\usepackage{multirow}
\usepackage{rotating}[figuresright]
\makeatletter

\newcommand{\Rmnum}[1]{\expandafter\@slowromancap\romannumeral #1@}
\makeatother
\usepackage{hyperref}
\usepackage{xcolor}
\begin{document}

\title{Evolution of the Three Spectral Components in the Prompt Emission of GRB 240825A}


\correspondingauthor{Shao-Lin Xiong, Rahim Moradi}
\email{xiongsl@ihep.ac.cn, rmoradi@ihep.ac.cn}

\author{Chen-Wei Wang}
\affil{Key Laboratory of Particle Astrophysics, Institute of High Energy Physics, Chinese Academy of Sciences, Beijing 100049, China}
\affil{University of Chinese Academy of Sciences, Chinese Academy of Sciences, Beijing 100049, China}

\author{Wen-Jun Tan}
\affil{Key Laboratory of Particle Astrophysics, Institute of High Energy Physics, Chinese Academy of Sciences, Beijing 100049, China}
\affil{University of Chinese Academy of Sciences, Chinese Academy of Sciences, Beijing 100049, China}

\author{Shao-Lin Xiong}
\affil{Key Laboratory of Particle Astrophysics, Institute of High Energy Physics, Chinese Academy of Sciences, Beijing 100049, China}

\author{Rahim Moradi}
\affil{Key Laboratory of Particle Astrophysics, Institute of High Energy Physics, Chinese Academy of Sciences, Beijing 100049, China}

\author{Yan-Qiu Zhang}
\affil{Key Laboratory of Particle Astrophysics, Institute of High Energy Physics, Chinese Academy of Sciences, Beijing 100049, China}
\affil{University of Chinese Academy of Sciences, Chinese Academy of Sciences, Beijing 100049, China}

\author{Chao Zheng}
\affil{Key Laboratory of Particle Astrophysics, Institute of High Energy Physics, Chinese Academy of Sciences, Beijing 100049, China}
\affil{University of Chinese Academy of Sciences, Chinese Academy of Sciences, Beijing 100049, China}

\author{Bing Li}
\affil{Key Laboratory of Particle Astrophysics, Institute of High Energy Physics, Chinese Academy of Sciences, Beijing 100049, China}

\author{Xiao-Bo Li}
\affil{Key Laboratory of Particle Astrophysics, Institute of High Energy Physics, Chinese Academy of Sciences, Beijing 100049, China}

\author{Cheng-Kui Li}
\affil{Key Laboratory of Particle Astrophysics, Institute of High Energy Physics, Chinese Academy of Sciences, Beijing 100049, China}

\author{Jia-Cong Liu}
\affil{Key Laboratory of Particle Astrophysics, Institute of High Energy Physics, Chinese Academy of Sciences, Beijing 100049, China}
\affil{University of Chinese Academy of Sciences, Chinese Academy of Sciences, Beijing 100049, China}

\author{Yue Wang}
\affil{Key Laboratory of Particle Astrophysics, Institute of High Energy Physics, Chinese Academy of Sciences, Beijing 100049, China}
\affil{University of Chinese Academy of Sciences, Chinese Academy of Sciences, Beijing 100049, China}

\author{Bo-Bing Wu}
\affil{Key Laboratory of Particle Astrophysics, Institute of High Energy Physics, Chinese Academy of Sciences, Beijing 100049, China}

\author{Sheng-Lun Xie}
\affil{Key Laboratory of Particle Astrophysics, Institute of High Energy Physics, Chinese Academy of Sciences, Beijing 100049, China}
\affil{Institute of Astrophysics, Central China Normal University, Wuhan 430079, China}

\author{Wang-Chen Xue}
\affil{Key Laboratory of Particle Astrophysics, Institute of High Energy Physics, Chinese Academy of Sciences, Beijing 100049, China}
\affil{University of Chinese Academy of Sciences, Chinese Academy of Sciences, Beijing 100049, China}

\author{Shu-Xu Yi}
\affil{Key Laboratory of Particle Astrophysics, Institute of High Energy Physics, Chinese Academy of Sciences, Beijing 100049, China}

\author{Zheng-Hang Yu}
\affil{Key Laboratory of Particle Astrophysics, Institute of High Energy Physics, Chinese Academy of Sciences, Beijing 100049, China}
\affil{University of Chinese Academy of Sciences, Chinese Academy of Sciences, Beijing 100049, China}

\author{Peng Zhang}
\affil{Key Laboratory of Particle Astrophysics, Institute of High Energy Physics, Chinese Academy of Sciences, Beijing 100049, China}
\affil{College of Electronic and Information Engineering, Tongji University, Shanghai 201804, China}

\author{Shuang-Nan Zhang}
\affil{Key Laboratory of Particle Astrophysics, Institute of High Energy Physics, Chinese Academy of Sciences, Beijing 100049, China}
\affil{University of Chinese Academy of Sciences, Chinese Academy of Sciences, Beijing 100049, China}

\author{Wen-Long Zhang}
\affil{Key Laboratory of Particle Astrophysics, Institute of High Energy Physics, Chinese Academy of Sciences, Beijing 100049, China}
\affil{School of Physics and Physical Engineering, Qufu Normal University, Qufu, Shandong 273165, China}

\author{Zhen Zhang}
\affil{Key Laboratory of Particle Astrophysics, Institute of High Energy Physics, Chinese Academy of Sciences, Beijing 100049, China}

\begin{abstract}
The prompt emission of Gamma-Ray Bursts (GRBs) could be composed of different spectral components, such as a dominant non-thermal Band component in the keV-MeV range, a subdominant quasi-thermal component, and an additional hard non-thermal component extending into the GeV range.
The existence and evolutionary behaviors of these components could place strong implication on physical models, such as ejecta composition and dissipation processes.
Although numerous GRBs have been found to exhibit one or two spectral components, reports of GRBs containing all three components remain rare.
In this letter, based on the joint observations of GRB 240825A from multiple gamma-ray telescopes, we conduct a comprehensive temporal and spectral analysis to identify the presence and evolution of all three components.
The bulk Lorentz factor of this bright and relatively short-duration burst is independently calculated using the thermal and hard non-thermal components, supporting a jet penetration scenario.
The multi-segment broken power-law feature observed in the flux light curves suggests the presence of an early afterglow in the keV–MeV band and hints at a possible two-jet structure.
Furthermore, the observed transition from positive to negative on the spectral lag can be interpreted as an independent evolution of the soft and hard components, leading to misalignment in the cross-correlation function (CCF) analysis of pulses.

\end{abstract}

\keywords{Gamma-ray burst}

\section{Introduction} \label{sec:intr}
Gamma-ray bursts (GRBs) manifest as a rapid increase in gamma-ray flux in the sky (the prompt emission), accompanied by a gradual and sustained change in flux in a wide wavelength from the radio band to the GeV band over the following hours or days (the afterglow) \citep{2018GRBbook}. 
There are three elemental spectral components in the prompt emission of GRB, including a non-thermal component usually described with smoothly joint broken powerlaw, a quasi-thermal component, and another non-thermal component extending to high energies (GeV to TeV band) \citep{overview_Zhang_2011}. 

For most GRBs, prompt emission is dominated by a smoothly joint broken powerlaw component, also known as the BAND component \citep{GRBM_Band_2005}, or a Cutoff Powerlaw (CPL) component in some cases. Although the physical origin of the phenomenological Band component remains under debate, it is currently widely interpreted as synchrotron radiation from non-thermal electrons, which is accelerated by the internal shocks or magnetic reconnection. 
Statistical studies of the low energy spectral index ($\alpha$) of the BAND component suggest an average value of $\sim1$ \citep{BATSEcatalog_Kaneko_2006,GBMcatalog_Poolakkil_2021,GBMcatalog_Burgess_2019}. 
However, since the synchrotron cooling time scale is typically much shorter than the time scale of GRBs, $\alpha$ is expected to be $-3/2$ in the ``fast cooling” regime, which is much softer than the observation\citep{fastcooling_Ghisellini_2000,fastcooling_Uhm_2014,2018GRBbook}. 
Additionally, many bursts have an $\alpha$ even harder than -2/3, which is the death line of the basic synchrotron model, known as the ``synchrotron death line" problem \citep{deathline_Preece_1998,deathline_Lloyd_2000,th_Meszaros_2000,deathline_Medvedev_2006,2018GRBbook}. 
Many studies have tried to tackle these problems, including the interpretation of this BAND component as modified thermal emission from a photosphere emission to solve the fast cooling problem and the synchrotron death line problem \citep{dissipative_photosphere_Rees_2005,dissipative_photosphere_Bhattacharya_2020,probability_photosphere_peer_2008,probability_photosphere_peer_2011,photosphere_Meng_2011}. 

Compared with the widely detected non-thermal BAND component, the (quasi-)thermal component is only observed in a small number of GRBs, even as the number of observed GRBs continues to increase. \citep{th_Burgess_2014,gamma_Asaf_2015,th_Ryde_2005}. This component is usually subdominant \citep{th_sample_2011_Guiriec,th_sample_2012_zhang,th_sample_2013_Guiriec}, yet the fireball model naturally predicts that a quasi-thermal component from photospheric emission should be superimposed on the non-thermal component \citep{th_Goodman_1986,th_Paczynski_1986,overview_Meszaros_2006,th_Ryde_2005,th_Meszaros_2000,rph_Daigne_2002}. Incorporating a thermal component into the spectral model often alleviates the ``synchrotron death line" issue (though not always) and improves the spectral fitting results of GRBs with $\alpha>-2/3$. Furthermore, extremely rare GRBs are found to be dominated by the thermal component, with GRB 090902B being a well-known example \citep{090902B_Abdo_2009,090902B_Ryde_2010,090902B_Zhang_2011}.

The high energy component (referred as the hard non-thermal component) is relatively independent and dominates in the GeV band, extending even into the TeV range \citep{09A_LHAASO_1,09A_LHAASO_2}. 
The origin of high-energy components, whether from the internal shock region or from the external shock region, remains a subject of hot debate.
For TeV emission, the observational features \citep{09A_LHAASO_1,line_2024_Zhang,KeVTeV_2024_Zhang} strongly indicate an external shock origin. However, the situation is more complex for GeV emission. 
First, it is important to note that GeV emission does not always constitute an independent spectral component. Some GRBs exhibit a spectrum that can be well described by a single BAND model extending from keV to GeV, as seen in GRB 080916C \citep{080916C_2009_Abdo}. However, in many GRBs with GeV emission, an additional high-energy component is observed, typically manifesting as a power-law spectrum in the GeV range. Some events also exhibit a high-energy cutoff \citep{090926A_2011_Ackermann,14C_Ajello_2020}. 
Since the first detection of GeV radiation from GRBs, the debate on its origin, whether from the internal or external shock region, has persisted.
Some research suggest an external origin, as the GeV emission follows a powerlaw temporal decay \citep{GeVEx_2010_Kumar,GeVEx_2010_Ghisellini,GeVEx_2024_Fraija}. 
However, certain GRBs exhibit characteristics that point to an internal origin, at least for the prompt GeV emission. These include rapid variability in the light curve, spectral continuity from keV to MeV (as in GRB 080916C), and a correlation between the high-energy and low-energy components \citep{090902B_Zhang_2011}.
An interesting observational phenomenon is the delayed onset of high-energy components relative to the BAND component, which has been widely reported \citep{080916C_2009_Abdo,090926A_2011_Ackermann,090902B_Abdo_2009,090902B_Zhang_2011,LATcatalog_2019_Ajello,14C_Ajello_2020}. 
Several interpretations have been proposed to explain this delay, including spectral hardening, the evolution of particle acceleration, and the changes in opacity.
\citep{090926A_2011_Ackermann,090902B_Zhang_2011}

Usually, only one or two spectral components can be distinguished in the prompt emission of GRBs. Events exhibiting all three spectral components are relatively rare. 
Here, we report the discovery and measurement of the three spectral components in the prompt emission of GRB 240825A, utilizing data from GECAM-B, GECAM-C, \textit{Insight}-HXMT, \textit{Fermi}/GBM and \textit{Fermi}/LAT. 

This paper is structured as follows: Observational details and data reduction are described in Section 2. 
In Section 3, we present the results of the temporal analysis and the spectral analysis, along with discussions on the evolution of the three spectral components and the initial bulk Lorentz factor. 
Finally, we summarize our findings in Section 4.

\section{Observation and Data reduction} \label{sec:redu}
\subsection{General Observation}
At 15:53:00.085 UT on 25 August 2024 (denoted as $T_0$), \textit{Fermi}/GBM \citep{GCN_25A_GBM} was triggered by this bright event, GRB 240825A, which is also observed by GECAM-B \citep{GCN_25A_GECAM}, GECAM-C, \textit{Insight}-HXMT and \textit{Swift}/BAT \citep{GCN_25A_BAT}. The prompt emission of GRB 240825A shows a profile with many rapid varing spikes superimposed on a bright narrow but smooth peak. The high energy emission of GRB 240825A is observed by \textit{Fermi} Large Area Telescope (LAT) \citep{GCN_25A_LAT}. The redshift of GRB 240825A is measured by subsequent follow-up observation as 0.659 \citep{GCN_25A_VLT}. The afterglow exhibits the typical behavior of type II GRBs, although no supernova has been reported until now \citep{GCN_25A_Ruffini,25A_Cheng_2024}.

\subsection{GECAM observation}
The GECAM constellation is composed of four instruments: GECAM-A/B \citep{GEC_INS_Li2022}, GECAM-C \citep{HEBS_INS_Zhang2023}, and GECAM-D \citep{GTM_INS_wang2024}, dedicated to monitor all-sky gamma-ray transients. As the first pair of micro-satllties, GECAM-A and GECAM-B are launched in December 10, 2020. GECAM-C is the third GECAM payload, which is on board SATech-01 \citep{SATech_01} and was launched in July 2022. 25 GRDs are equipped on GECAM-B while 12 GRDs are equipped on GECAM-C. All 25 GRDs of GECAM-B and 10 out of 12 GRDs of GECAM-C operate in two readout channels: high gain (HG) and low gain (LG) \citep{GEC_INS_An2022,HEBS_INS_Zhang2023}. For spectral analysis, the energy range is 40-300\,keV for HG and 0.7-6\,MeV for LG. For temporal analysis, the energy range is 40-300\,keV for HG and 0.3-6\,MeV for LG. Detectors with an incident angle less than 60$^\circ$ (that is, GRD21 and GRD22 of GECAM-B for this burst) are selected for spectral and temporal analysis with GECAMTools \footnote{https://gecam.ihep.ac.cn/grbDataAnalysisSoftware.jhtml}. The lightcurve of GECAM-B and GECAM-C are shown in Figure\,\ref{fig:lc}a. Since the incident angle of GECAM-C GRD01 (which has the smallest incident angle among all GECAM-C GRDs during this burst) is greater than 60$^\circ$. GECAM-C data are not used in the spectral analysis.

\subsection{\textit{Insight}-HXMT observation}
\textit{Insight}-HXMT is China's first X-ray astronomy satellite, which consists of three collimated telescopes, namely the high energy X-ray telescope (HE), the medium energy X-ray telescope (ME) and the low energy X-ray telescope (LE) \citep{HXMT_zhang_2018,HXMT_zhang_2020,HXMT_Li_2020}. HE consists of 18 NaI(Tl)/CsI(Na) phoswich scintillation detectors with collimator. For GRB detection, gamma-ray can penetrate the satellite platform and be detected by CsI detectors of HE with a large effective area \citep{HE_calibration}. And the energy range of CsI data used for spectral analysis is 120 to 600\,keV, while for temporal analysis the energy range is 30 to 1000\,keV. The lightcurve of all the CsI detectors are shown in Figure\,\ref{fig:lc}a. The HXMT data is only used for fitting the time-integrated spectrum rather than the time-resolve spectra analysis. 

\subsection{Fermi observation}

\begin{figure*}
\centering
\begin{tabular}{cc}
\begin{minipage}[b]{0.5\linewidth}
    \begin{overpic}[width=\textwidth]{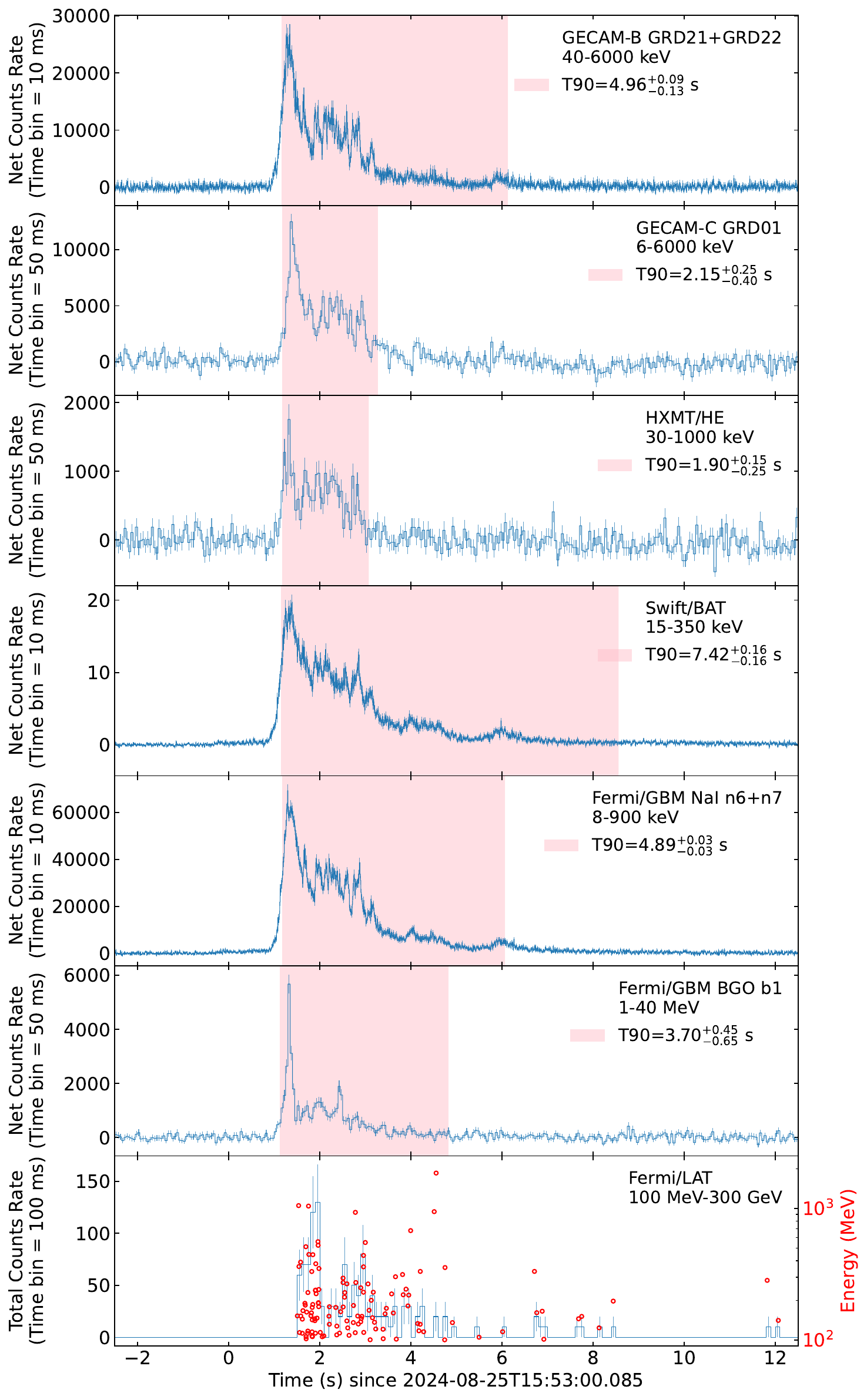}\put(0, 100){\bf a}\end{overpic}
\end{minipage}
\begin{minipage}[b]{0.4\linewidth}
        \begin{overpic}[width=\textwidth]{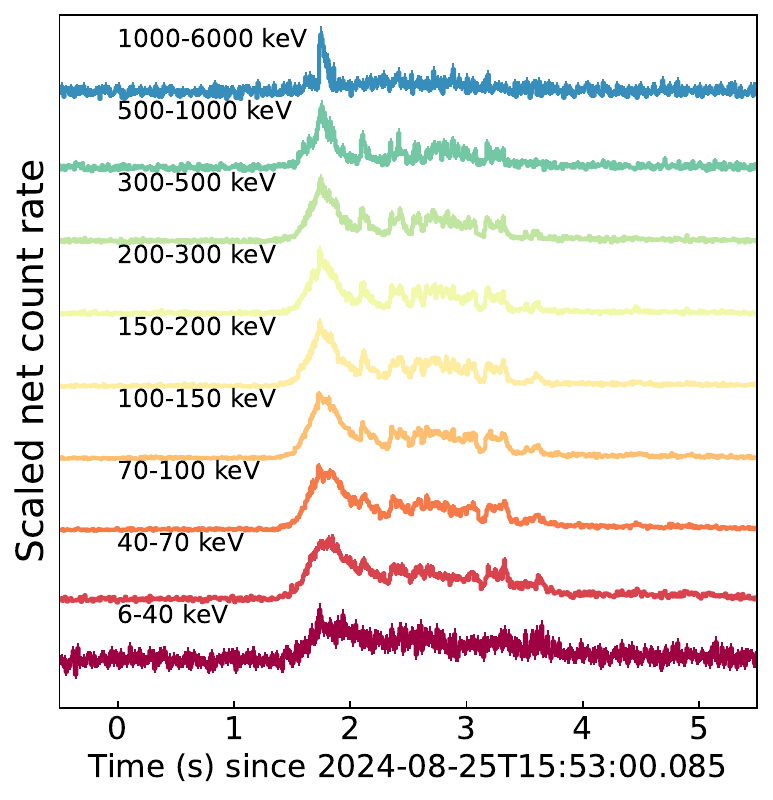}\put(-4, 90){\bf b}\end{overpic}
        \begin{overpic}[width=\textwidth]{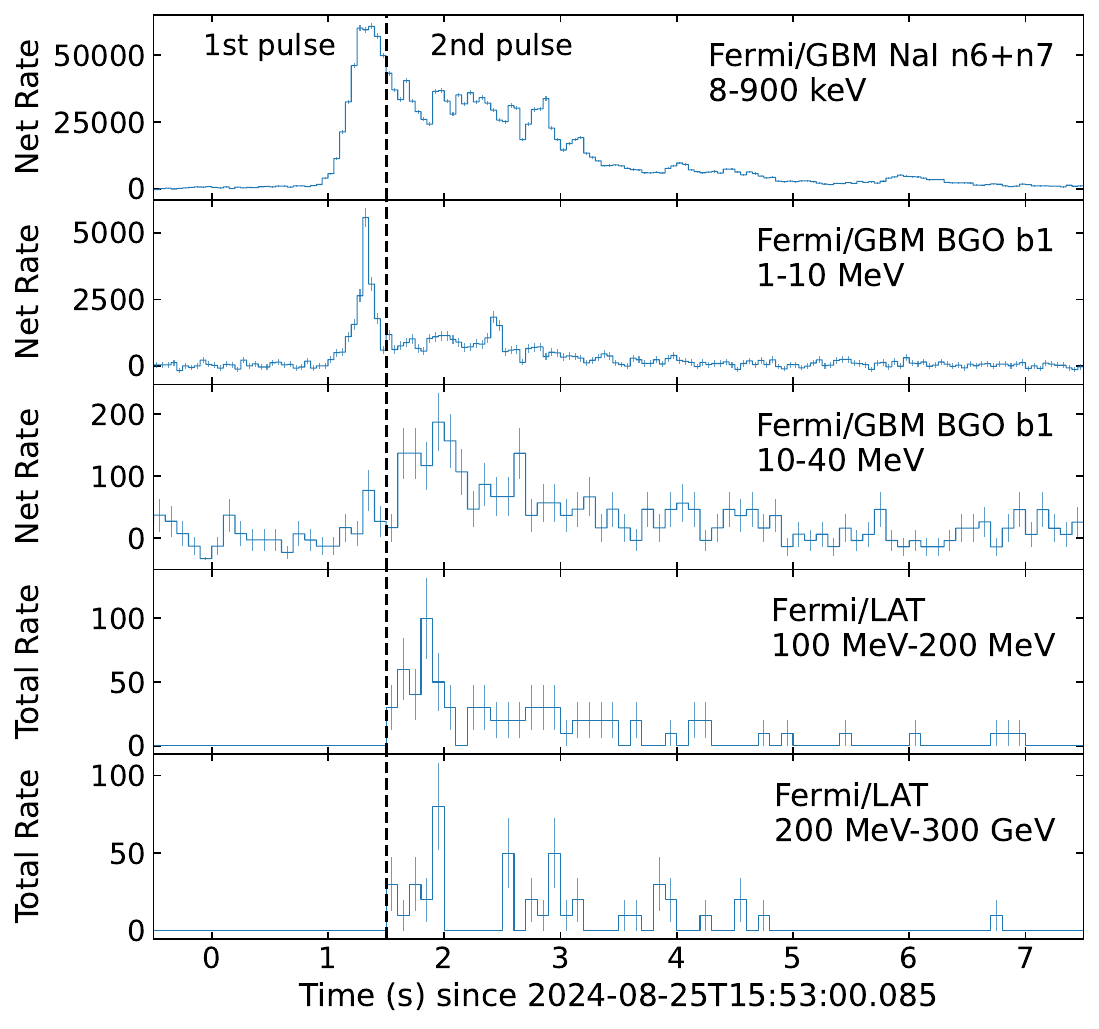}\put(-4, 90){\bf c}\end{overpic}
\end{minipage} \\
\multicolumn{2}{c}{\begin{overpic}[width=0.9\textwidth]{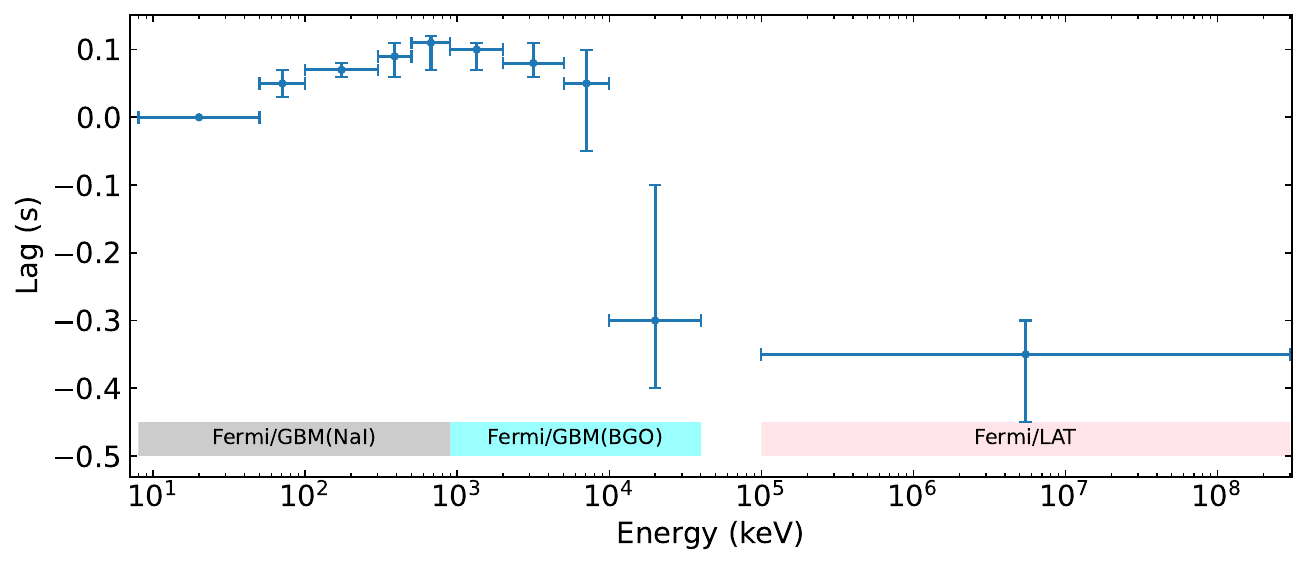}\put(-5, 40){\bf d}\end{overpic}} \\
\end{tabular}
\caption{\textbf{The lightcurve of GRB 240825A.} (a), the lightcurve of GRB 240825A observed by different mission, with the corresponding $T_{90}$. 
(b), the lightcurve of different energy range from keV to MeV observed by GECAM-B. It can be seen that the main pulse is actually consist of a soft, slow component, as well as a hard and narrow component. 
(c), the lightcurve of different energy range from keV to GeV. The lightcurves of BGO and LAT are divided into two energy bands based on the dominant spectral components. The profile of the lightcurve in 10-40 MeV exhibits a bump shape, corresponding to the high-energy non-thermal component in the spectrum, which is significantly different from the lightcurve below 10 MeV. 
(d), the spectral lag of GRB 240825A from keV to GeV, which shows a ``positive to negative" behavior.
All error bars on data points represent their 1$\sigma$ confidence level. 
}
\label{fig:lc}
\end{figure*}

The Gamma-ray Burst Monitor (GBM) is one of the two instruments onboard the \textit{Fermi} Gamma-ray Space Telescope \citep{GBM_meegan_09,GBM_Bissaldi_09}. \textit{Fermi}/GBM is composed of 14 detectors with different orientations: 12 Sodium Iodide (NaI) detectors (labeled from n0 to nb).  For spectral analysis, the energy range is 8-900\,keV for NaI detector and 0.3-40\,MeV for BGO detector. The NaI detectors with incident angle less than 60$^\circ$ (i.e. n6 and n7) are selected for analysis and the BGO detector b1 is utilized for analysis. For temporal analysis, the energy range for NaI detector is 8-900\,keV and for BGO detector is 1-40\,MeV (Figure\,\ref{fig:lc}a). Data are reduced by GBM data tools \footnote{https://fermi.gsfc.nasa.gov/ssc/data/analysis/gbm/gbm\_data\_tools/gdt-docs/}.

The Large Area Telescope (LAT) is the other instrument onboard the \textit{Fermi} Gamma-ray Space Telescope, which is a pair-conversion telescope comprising a 4 × 4 array of silicon strip trackers and cesium iodide (CsI) calorimeters \citep{LAT_Atwood_09}. We selected P8R3\_TRANSIENT020E\_V3 class events in the 100 MeV–300 GeV energy range from a region of interest (ROI) of 12$^{\circ}$ radius centered on the burst location (Figure\,\ref{fig:lc}a). The unbinned analysis method is used for LAT photons from 0.1 to 300 GeV by Fermitools \citep{Fermitools}. As for the background, the contributions from Galactic diffuse and isotropic emission (assumed to be a powerlaw spectrum), as well as the contributions from the Fermi Point Source catalog (assumed to be a powerlaw2 spectrum) are taken into consideration. 

\subsection{Swift observation}
The Burst Alert Telescope (BAT) is a coded aperture imaging instrument onboard the Swift. The data were downloaded from the UK Swift Science Data Center and extracted by High Energy Astrophysics software package (HEAsoft) \citep{HEAsoft} v.6.32.1. For temporal analysis, the energy range is 15-350\,keV (Figure\,\ref{fig:lc}a) and for spectral analysis, the energy range is 15-150\,keV.

\subsection{Spectral analysis}

Spectral fitting is performed using PyXspec software \citep{pyxspec}. 
Three models, CPL, BAND and Blackbody (expressed as bbodyrad) and their combination are used in spectral fitting. 
The CPL model is expressed as
\begin{equation}   
N(E)=A\left(\frac{E}{E_0}\right)^{\alpha} {\rm exp}(-\frac{E}{E_{\rm c}}),
\label{equ:CPL_Model}
\end{equation}
where $A$ is the normalization constant ($\rm photons \cdot cm^{-2} \cdot s^{-1} \cdot keV^{-1}$), $\alpha$ is the power law photon index, $E_0$ is the pivot energy fixed at 1 keV, and $E_{\rm c}$ is the characteristic cutoff energy in keV. The peak energy $E_{\rm p}$ is related to $E_{\rm c}$ through $E_{\rm p}$=$(2 + \alpha)E_{\rm c}$.
The Band model is expressed as
\begin{equation}
N(E)=\left\{
\begin{array}{l}
A(\frac{E}{100\,{\rm keV}})^{\alpha}{\rm exp}(-\frac{E}{E_{\rm c}}),\,E<(\alpha-\beta)E_{\rm c}, \\
A\big[\frac{(\alpha-\beta)E_{\rm c}}{100\,{\rm keV}}\big]^{\alpha-\beta}{\rm exp}(\beta-\alpha)\\\,\,\,\,\,(\frac{E}{100\,{\rm keV}})^{\beta}, E\geq(\alpha-\beta)E_{\rm c}, 
\end{array}\right.
\label{equ:band_Model}
\end{equation}
where $A$ is the normalization constant ($\rm photons \cdot cm^{-2} \cdot s^{-1} \cdot keV^{-1}$), $\alpha$ and $\beta$ are the low-energy and high-energy power law spectral indices, $E_{\rm c}$ is the characteristic energy in keV and the peak energy $E_{\rm p}$ is related to $E_{\rm c}$ through $E_{\rm p}$=$(2 + \alpha)E_{\rm c}$.
The bbodyrad model is expressed as
\begin{equation}   
N(E)=\frac{1.0344 \times 10^{-3}\times AE^2}{{\rm exp}(\frac{E}{kT})-1},
\label{equ:bb_Model}
\end{equation}
where $A=R^2_{\rm km}/D^2_{10}$ is the normalization constant, $R_{km}$ is the source radius in km and $D_{10}$ is the distance to the source in units of 10 kpc, and $kT$ is the temperature in keV.

The Bayesian Information Criterion (BIC) is used for both the model comparison of spectral analysis and temporal analysis, which is defined as BIC$=-2\ln L+k\ln N$, where L represents the maximum likelihood value, $k$ denotes the number of free parameters in the model, and $N$ signifies the number of data points. For each model, a BIC value is calculated and a minimal BIC value, BIC$\rm_{min}$ is obtained. If only one well constrained model's BIC falls in the range of BIC$\rm_{min}$ to BIC$\rm_{min}$+2, this model will be selected as the best model. But when two or more models' BIC fall in the range of BIC$\rm_{min}$ to BIC$\rm_{min}$+2, if a (combining) model contains a BAND component and the parameters are well constrained, the (combining) model is selected as the best model; otherwise, we choose the (combining) model with the lowest BIC as the best model.

\section{Analysis Results and Discussion} \label{sec:diss}
The analysis includes aspects of temporal and spectral. 
In this section, we will first discuss the detailed spectral analysis result of GRB 240825A, which clearly shows the evolution of three components. Then we provide the constraint of the bulk Lorentz factor ($\Gamma$) based on two independent methods. Some other interesting features will also be discussed. 
The trigger time $T_0$ is assumed as the time zero point of multi-segment broken powerlaw in all temporal analyses.

\subsection{Three components of the spectrum} 
\begin{figure*}
\centering
\begin{tabular}{cc}
\begin{overpic}[width=0.45\textwidth]{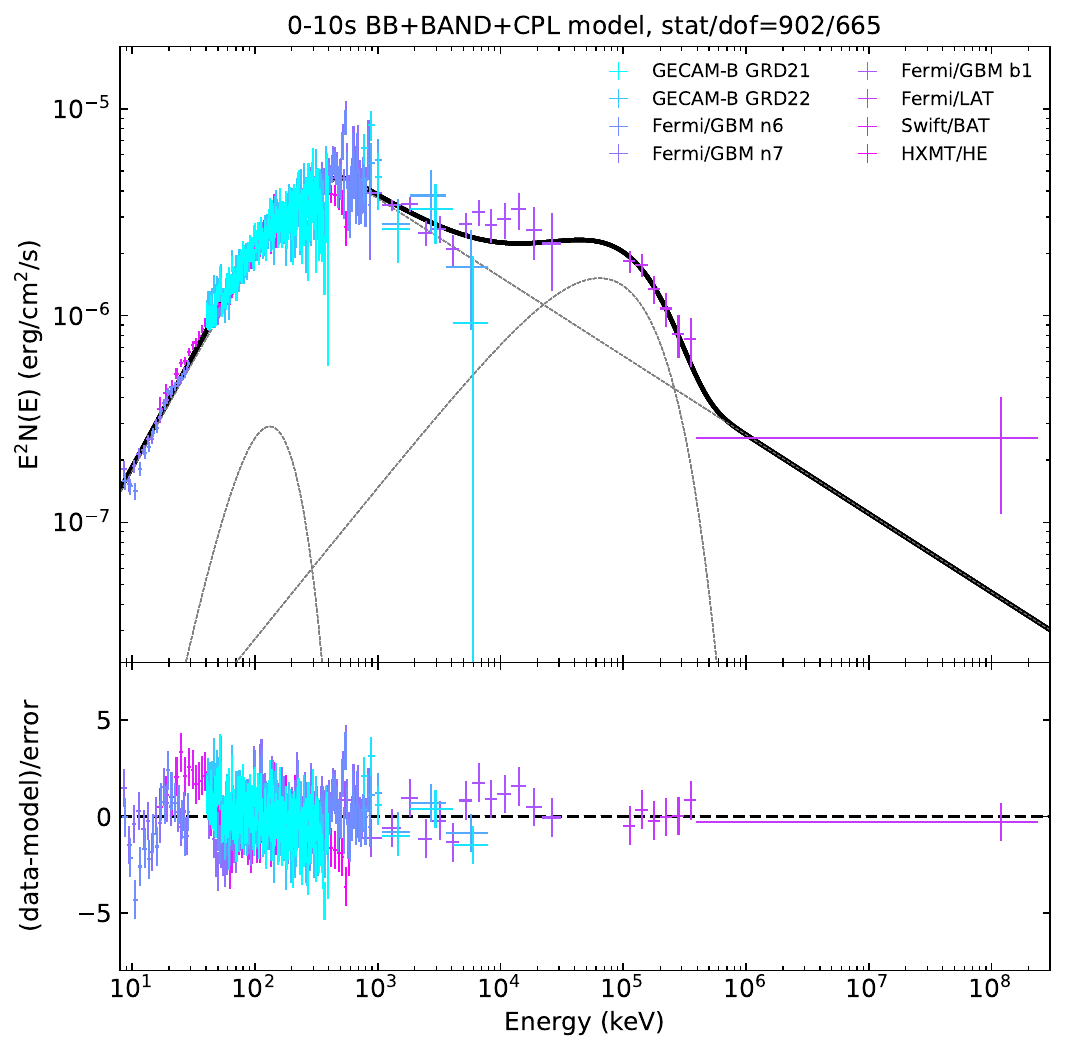}\put(0, 90){\bf a}\end{overpic} &
        \begin{overpic}[width=0.45\textwidth]{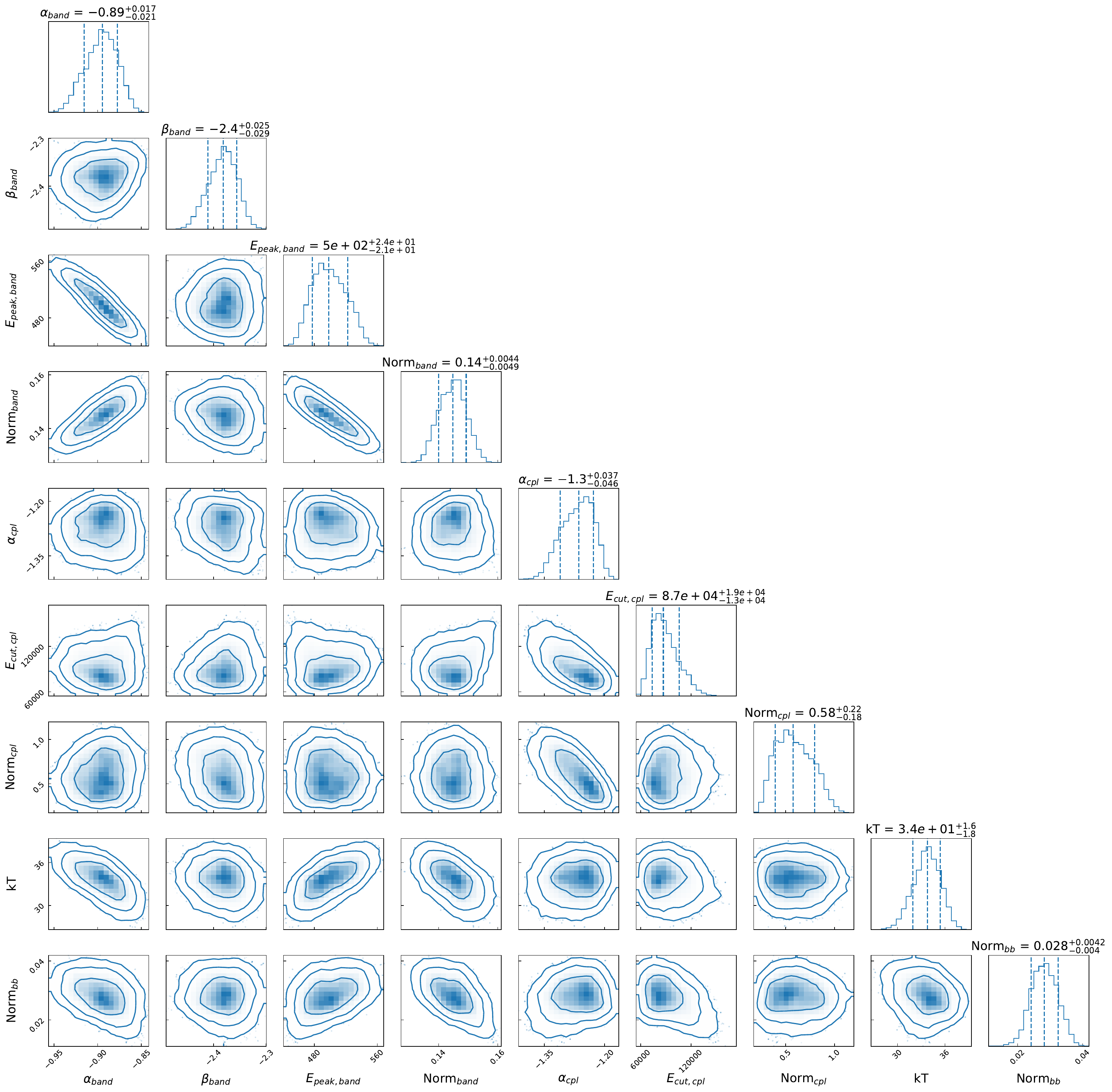}\put(0, 90){\bf b}\end{overpic} \\
\begin{overpic}[width=0.45\textwidth]{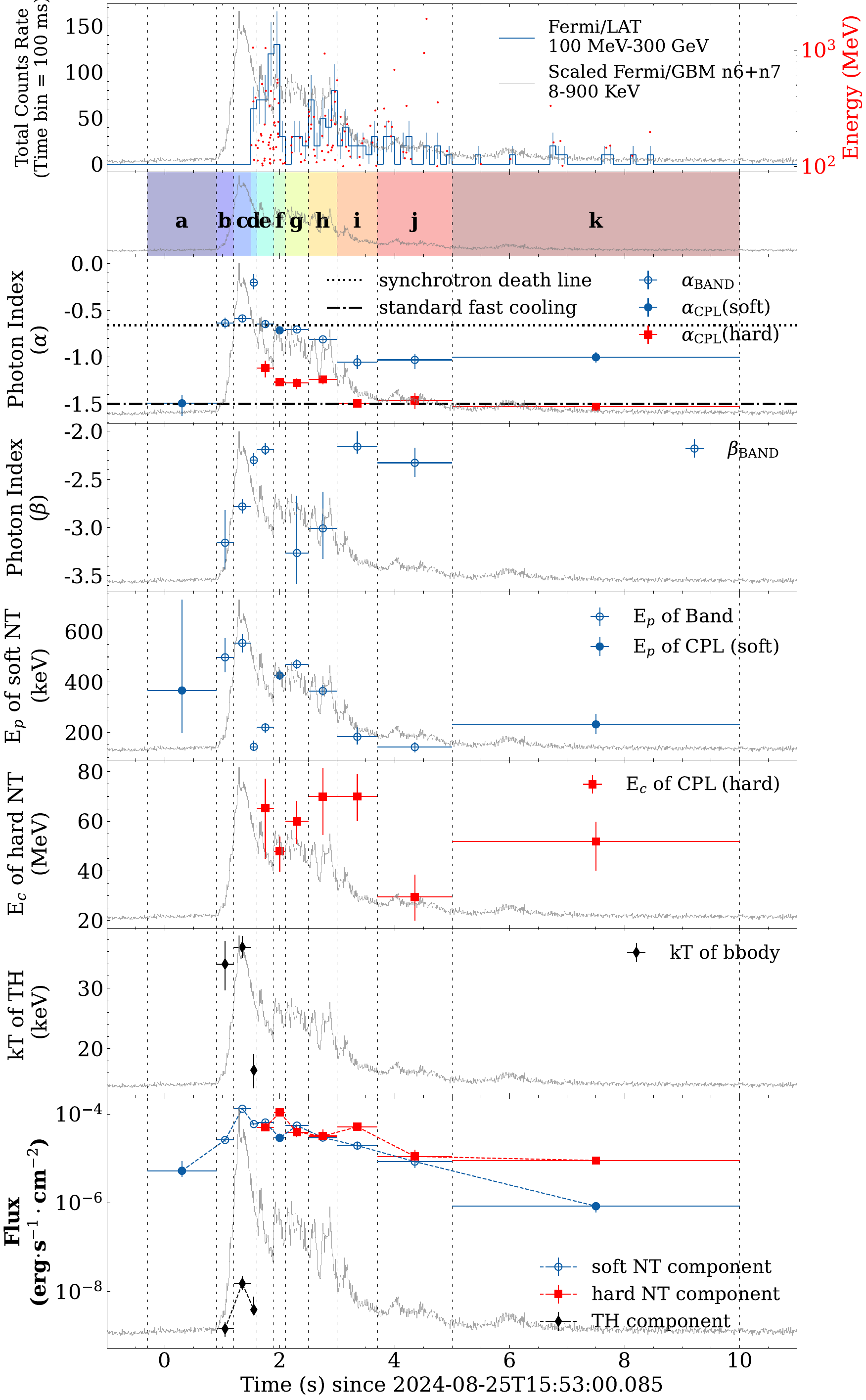}\put(-2, 90){\bf c}\end{overpic} &
        \begin{overpic}[width=0.42\textwidth]{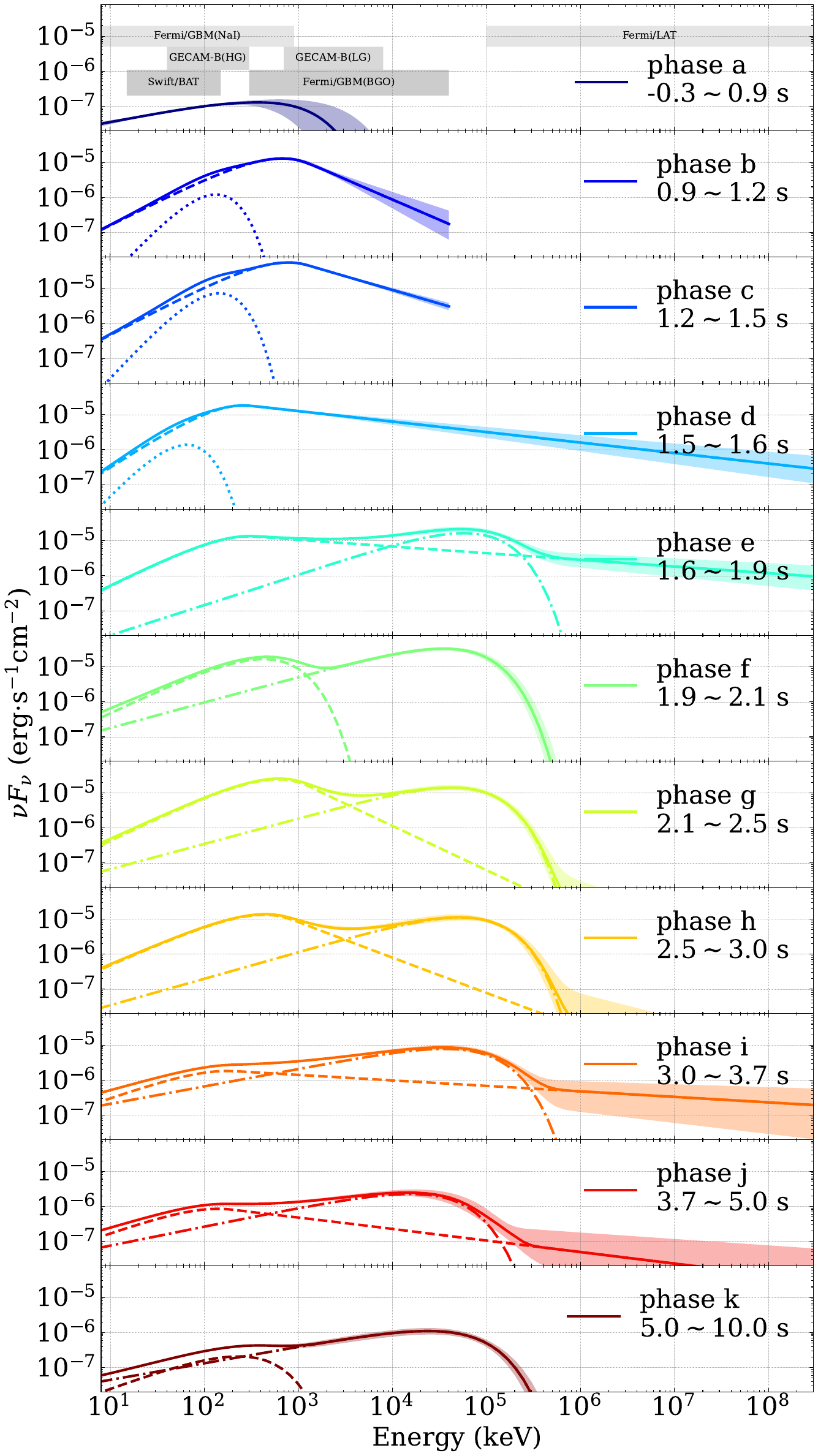}\put(-2, 90){\bf d}\end{overpic} \\
\end{tabular}
\caption{\textbf{The time integrated and time resolved spectra of GRB 240825A.} (a)(b), the time integrated spectrum (S-I) of GRB 240825A, including a soft Non-thermal component in BAND shape, a hard Non-thermal component in CPL shape, and a thermal component. The flux is calculated in energy range from 10\,keV to 100\,GeV. 
(c)(d), the time resolved spectra (S-II) of GRB 240825A. The thermal component is only exist in phase \textit{c},\textit{d} and \textit{e}, during which the hard non-thermal component (GeV emission) has not yet appeared. 
All error bars on data points represent their 1$\sigma$ confidence level. All shaded areas around the best-fit lines represent their 1$\sigma$ confidence bands.
}
\label{fig:spec}
\end{figure*}

A notable feature of GRB 240825A is the bright GeV emission, manifested as a hump structure (descried by a CPL model) superimposed on the high-energy extension of the low-energy BAND spectrum in the time integrated spectrum (denoted as S-I), as shown in the Figure\,\ref{fig:spec}a and Table\,\ref{tab:spec_fit}. To investigate the possible photosphere emission, we tried adding a thermal component in the fitting of S-I, and the BAND+CPL+BB shows a much lower BIC than the result of BAND+CPL as well as a better stat/dof (Table\,\ref{tab:spec_fit}), all the parameters are constrained well (Figure\,\ref{fig:spec}b).

\subsubsection{Early thermal component}
To confirm whether this thermal component could be a spurious structure generated by spectral evolution, we divided 11 time resolved spectra (denoted as S-II) based on the profile of the lightcurve, labeled as phase \textit{a} to phase \textit{k} (Figure\,\ref{fig:spec}c). Phase \textit{a} covers the weak emission before the main peak. Phase \textit{b} and \textit{c} cover the time intervals of the main peak before the GeV emission appears. Phase \textit{d} has a narrow time width, covering the time interval from the appearance of GeV emission to the end of the smoothly decay part of the main peak (where some new pulses start to become prominent). Phase \textit{e} to phase \textit{h} divide the dense narrow pulses on the decay edge of the main peak. Phase \textit{i} to phase \textit{k} cover the remaining part of the main pulse while ensuring sufficient counts to fit the SED from keV to GeV. For each spectrum of S-II, seven models are used for fitting, including single CPL, single BAND, BAND+CPL, CPL+CPL, BAND+BAND, BB+BAND, BB+CPL, as listed in Table\,\ref{tab:spec_fit}. 

The result of S-II suggests that BB+BAND is more favored during phase \textit{b}-\textit{d}, and indicates that temperature (kT) of the thermal components evolves with flux. Therefore, this thermal component is inclined to be real rather than a spurious structure caused by spectral evolution. 
We further studied the evolution of the thermal component by fitting time-resolved spectra with higher time resolution (denoted as S-III). S-III only covers the time interval when the presence of the blackbody is confirmed (i.e. phases \textit{b}, \textit{c}, and \textit{d} of S-II). Since there is no GeV emission in this period, the LAT data is excluded from the spectral fitting of S-III. The results of S-III further confirm the presence of the thermal component. Moreover, both the temperature and the flux (flux of the thermal component and non-thermal component in 10 to 1000 keV) follow a two-segment broken powerlaw evolution, and all the break times are consistent within the error, roughly at around the peak of lightcurve.

\subsubsection{Lower energy non-thermal component}
In most time intervals, the soft non-thermal component can be well described by the BAND model, while in few time intervals (phase \textit{a}, phase \textit{f} and phase \textit{k}), the CPL model is more applicable. Among them, phase \textit{a} and phase \textit{k} are in the initial and final stages of the prompt emission, while phase \textit{f} is in the middle stage of the burst. The BAND model is more favored in the time slices before and after phase \textit{f}, while a transition occurs in phase \textit{f}. If we consider the CPL model as a BAND model with a very soft high-energy photon index ($\beta$), it can be seen from Figure\,\ref{fig:spec}b that the $\beta$ of the soft non-thermal component undergoes a hardening process in the early stages of the burst, when the thermal components exist. In phases \textit{f}, \textit{g}, and \textit{h} (when the ``plateau-like" narrow pulse ensemble appears), $\beta$ rapidly softens, and then returns to the level before the ``plateau-like" appeared after the ``plateau-like" ends. 

The $E_{\rm p}$ evolution of the soft non-thermal component is also associated with the ``plateau-like" structure. $E_{\rm p}$ shows a ``intensity-tracking" pattern during the main pulse, and became harder during the ``plateau-like" and falls back to the same level as phase \textit{d} (the decay edge of the main pulse) after the ``plateau-like" ends. 

We also notice that if the thermal component is not included in the spectral model in phase \textit{b} and phase \textit{c}, not only is the BIC larger, but also the low-energy photon index ($\alpha$ in the BAND model) of the soft non-thermal component will exceed the synchrotron death line, and the inclusion of the thermal component can significantly improve this, but not complete solve this excess. Especially for phase \textit{d}, where even with the thermal component, the low-energy photon index still significantly exceeds the synchrotron death line (-2/3). This may suggest that the spectral index produced by non-thermal radiation itself is indeed very hard, exceeding the synchrotron death line. The low energy photon index for all time slice in S-II exceeds the prediction of standard fast-cooling of synchrotron (-3/2).

\subsubsection{Higher energy non-thermal component}
GeV emission, the hard non-thermal component, was detected approximately 1.5 seconds after $T_0$. 
During phase \textit{d}, the initial appearance of GeV emission, both GeV and MeV emission can be described by the same BAND model. 
Subsequently, an independent hard non-thermal component (the hard non-thermal component) becomes significant. 
From phase \textit{e} to phase \textit{k}, the $\alpha$ of the hard non-thermal component continues to soften, while the cutoff energy (E$\rm_c$) remains nearly constant within the error range.

It need to be note that, the soft non-thermal components have extended to the E$_c$ of the hard non-thermal components, not only in the time integral spectrum, but also in the time resolved spectral, which is also shown by \cite{25A_Zhang_2025}. 
If considering the $E_{\rm c}$ of the hard non-thermal components is produced the $\gamma\gamma$ annihilation, some cases can be discussed in such a scenario: 
\begin{enumerate}[(1)]
\item The hard non-thermal components and soft non-thermal components are from the same region, and the threshold photon energy of annihilation production is identical, then both the two components should follow the same $E_{\rm c}$.
\item The hard non-thermal components and soft non-thermal components are from different regions, and the hard non-thermal components have a larger radius, then the soft non-thermal component should exhibit a smaller $E_{\rm c}$ than the hard non-thermal components (at least can not extend beyond the $E_{\rm c}$ of hard non-thermal component), unless the jet undergoes significant deceleration within these few seconds.
\item The hard non-thermal components and soft non-thermal components are from different regions, and the hard non-thermal components have a smaller radius, it is natural for the hard non-thermal component to have a smaller $E_{\rm c}$ compared to the soft non-thermal component. However, this strongly contradicts the fact that the variability of the former is much longer than that of the latter. \cite{090902B_Zhang_2011} point that, if the jet is Poynting-flux dominated, such as the ICMART model, the emission radius is no longer associated with the time variability but with the time duration, and the emission radius is much larger than that in the internal model. For one hand, this allows a Band component to extend to very high energies, and for the other hand, the thermal components should be greatly suppressed in the Poynting-flux dominated jet and conflict with the thermal component observed in this burst. 
\end{enumerate} 
If the high-energy end of the hard non-thermal components follows a power-law decay (e.g. can be described by the BAND model) rather than an exponential cutoff (e.g. can be described by CPL), it could avoid these contradictions. However, we have found that for the majority of the time, the parameters of the BAND+BAND model cannot be constrained (the $\beta$ of the hard non-thermal component becomes softer than -10 and fails to converge, effectively reducing to an exponential cutoff). Only within a certain time segment (phase \textit{e}, which is also consistent with \cite{25A_Zhang_2025}) can it be properly fitted, yet it still does not represent the best model. 

On the other hand, a cutoff in the spectra can also be predicted by multiple inverse Compton scattering by a population of relativistic electrons with energy distribution of powerlaw spectrum \citep{Radiative_1986_Rybicki}. Based on the sources of the electron population and photon population, two scenarios are usually discussed: synchrotron self-Compton (SSC) and inverse-Compton (IC). 
In the scenario of SSC, the peak energy of SSC is given by $\nu^{\rm SSC}_{\rm peak}\approx2\gamma^2_e\nu^{\rm syn}_{\rm peak}$ for the Thomson regime, and $h\nu^{\rm SSC}_{\rm peak}\approx2\gamma_e \Gamma m_ec^2/(1+z)$ for the Klein-Nishina regime, where $\gamma_e$ = min($\gamma_c$,$\gamma_m$), $\gamma_m$ is the injection minimum energy for slow cooling and $\gamma_c$ is the cooling energy for fast cooling. Considering $E^{\rm SSC}_{\rm peak}\sim80\rm\,MeV$ and $E^{\rm syn}_{\rm peak}\sim0.5\rm\,MeV$ (S-I of Table\ref{tab:spec_fit}), a small $\gamma_e\lesssim10$ is implied in both Klein-Nishina and Thomson scatter regimes, which will lead to an extremely small efficiency of the electron acceleration and an extremely large magnetic field for synchrotron radiation \citep{14C_2019_MAGIC,KeVTeV_2024_Zhang}. 

As for the scenario of (multiple) IC, the seed photos are scattered to high energy in a region different from where it was produced and lead to a Comptonized spectrum. Considering that both the photons of BAND components and thermal components can play the role of seed photons, lengthy and complicated treatment has to be done to provide a specific and reasonable constraint of the IC theory, which is beyond the aim of this work. But since the soft BAND component covers a broader energy range than the hard CPL component in GRB 240825A, it is challenging to achieve a narrower distribution through scattering if the BAND components serve as the seed photons. Additionally, we note that it is interesting that the hard non-thermal components appear only after the thermal component has almost disappeared, which suggests a possible association between these two components and remains a subject for further research.

\subsection{Spectral lag}
We also calculated the spectral lag for any pair of light curves between the lowest energy band (8-50 keV) and any other nine energy bands (i.e. 50-100 keV, 100-300 keV, 300-500 keV, 500-900 keV, 0.9-2 MeV, 2-5 MeV, 5-10 MeV, 10-40 MeV and 0.1-300 GeV) using the CCF \citep{lag_liu_2022,lag_2010_Ukwatta,lag_2000_Norris}. The lightcurves with energy lower than 900 keV are extracted from the NaI detector of \textit{Fermi}/GBM, and the lightcurves in energy range of 0.1-300 GeV are extracted from \textit{Fermi}/LAT, while lightcurves of other energy bands are reduced from the BGO detector of \textit{Fermi}/GBM.

The spectral lag of GRB 240825A shows clearly a positive-to-negative transition, which is depicted in Figure\,\ref{fig:lc}d. 
When we take both the spectrum and the lightcurve into consideration, we can see that the hard non-thermal component is the dominant component above approximately 10 MeV while the profile of the lightcurve in 10-40 MeV is a single bump shape, corresponding to this hard non-thermal component in the spectrum, and the lightcurve lower than 10 keV is corresponding to the soft non-thermal component (described by the BAND model) with a bright pulse at $\sim$1\,s. 
As the BAND component, which dominated the spectrum below 10\,MeV, becomes the dominant component above 200\,MeV again. Thus, it is expected to see a similarity of the lightcurves between 0.2-300 GeV and lower than 10 MeV, both of which should roughly have two episodes. But the statistics of photons in 0.2-300 GeV are limited, so we cannot draw such a conclusion.

The results from the association between the spectrum and the lightcurve indicate that the positive lag originates from the spectral evolution of the BAND component, while the negative lag is associated with the onset of the GeV emission. 
Although both the positive lag between the keV-MeV band and the negative lag between the keV-GeV band are physical, the positive-to-negative transition, and also the break of the spectral lag at $\sim$1\,MeV are actually the combined effect of the independent evolution of the soft and hard components, in other words, the incorrect pulse alignment of the CCF. 
Hence this also provides a hint that when using spectral lags to test fundamental physics, such as Lorentz invariance, it should be treated with caution whether positive-to-negative transition of the spectral lag is of physical origin or an artifact caused by the limitation of analysis method.



\subsection{Constraint on the Lorentz factor $\Gamma$} \label{sec:gamma}
In the GRB fireball model, the fireball initially undergoes a rapid adiabatic acceleration from an initial radius of the fireball $r_0$. After the time when the outflow is accelerated to the maximum Lorentz factor, it will enter the coast phase with a constant Lorentz factor, during which a fraction of the kinetic energy is dissipated by internal shock during the prompt emission, resulting in a reduction of the Lorentz factor. 
Before the internal shock is developed, the fireball will first reach the photosphere radius $r_{\rm ph}$, which is defined as the radius above which the photon optical depth for Thomson scatter is below unity, with a Lorentz factor $\Gamma_{\rm ph}$ at the photosphere and produce (quasi-)thermal photosphere emission. 
Eventually, the fireball is decelerated by the circumburst medium (wind or ISM) with an initial Lorentz factor $\Gamma_0$. 
Considering the dissipation of the kinetic energy, it is expected that $\Gamma_0\lesssim \Gamma_{\rm ph}$ for an impulse energy injection \citep{gamma_Asaf_2007,gamma_2021_Zhang,2018GRBbook,14C_2023_Li}.

It has been shown that $\Gamma_{\rm ph}$ and the initial fireball radius $r_0$ can be revealed by the characteristic break in the temporal behavior of the thermal emission with in the matter-dominated fireball scenario \citep{gamma_Asaf_2007,gamma_Asaf_2015,2018GRBbook,gamma_2021_Zhang}. Before the break, there is no significant energy dissipation below the photosphere, while after the break, inner engine activity decreases, and the emission is dominated by the high-latitude effect. 
In such a scenario, the bulk Lorentz factor at the site of the photosphere $\Gamma_{\rm ph}$ can be described as 
\begin{equation}
    \Gamma_{\rm ph}=\left[ 1.06(1+z)^2\,D_L\frac{YF^{\rm ob}\sigma_{\rm T}}{2m_{\rm p}c^3R}\right]^{1/4}
    \label{equ:gamma_photosphere},
\end{equation}
and the photospheric radius is given by
\begin{equation}
    r_{\rm ph}=\frac{L\sigma_{\rm T}}{8\pi\Gamma_0^3m_{\rm p}c^3},
    \label{equ:rph_photosphere}
\end{equation}
and $r_0$ can be described as
\begin{equation}
    r_0=0.6\frac{D_L}{(1+z)^2}\left(\frac{F^{\rm ob}_{\rm BB}}{YF^{\rm ob}}\right)^{3/2}R,
    \label{equ:r0_photosphere}
\end{equation}
where $\sigma_{\rm T}$ is the Thomson's cross section, $R\equiv(F^{\rm ob}_{BB}/\sigma T^{4})^{1/2}$, $T$ is the temperature of the thermal component, $\sigma$ is Stefan–Boltzmann constant, $F^{\rm ob}_{\rm BB}$ is the flux of thermal component and $F^{\rm ob}$ of the total flux of thermal component and non-thermal component, $Y$ is the ratio between the total energy released and the energy observed ($Y\geq1$) \citep{gamma_Asaf_2007,rph_Meszaros_2002,rph_Daigne_2002}. 

As shown in the Figure\,\ref{fig:Gamma}a, both the temperature and the flux of the thermal component can be well described by a broken powerlaw, with a same break time with in the error. Before the break, the temperature is roughly constant (consistent within the error). At the break time, $T=45.14^{+5.32}_{-5.78}\,\rm keV$, $R=1.86^{+0.25}_{-0.22}\times10^{18}$, with Equation\,\ref{equ:gamma_photosphere} Equation\,\ref{equ:rph_photosphere}, and Equation\,\ref{equ:r0_photosphere}, $r_0=2.11^{+0.69}_{-0.65}\times10^8\,\rm cm$, $r_{\rm ph}=2.27^{+0.22}_{-0.23}\times10^{11}\,\rm cm$, and the $\Gamma_{\rm ph}=365^{+13}_{-11}$ can{\rm}be obtained with the assumption of $Y=1$. 
The result of the photoshere method is compared with other sample analyzed by \cite{gamma_Asaf_2015} with the same method is shown in Figure\,\ref{fig:Gamma}c. The results indicate that GRB 240825A falls within the typical position of the samples. 

A particular aspect is that, after break occurs, the evolution of the thermal component is very intense with large powerlaw index, which is much larger than the slope observed in most GRBs \citep{gamma_Asaf_2007}, indicating that the dissipation process is very quick. 
We also note that this result is based on the assumption of a pure fireball scenario. If the jet is hybrid with a hot fireball component and a cold Poynting-flux component, the result should be modified as the treatment by \cite{gamma_Gao_2015}.

Additionally, since many hard $\gamma$ photons with energy much higher than 511\,keV are detected in GRB 240825A with a large $E_{\rm iso}=1.87^{+0.01}_{-0.01}\times10^{53} \rm\,erg$, a large $\Gamma_0$ is needed to avoid the ``compactness problem", otherwise the high opacity originating from $\gamma\gamma$-annihilation will prevent the escape of MeV-GeV photons \citep{compactness_Baring_1997}. 

Assuming that the $E_{\rm c}$ of the high energy components results from the $\gamma\gamma$-annihilation, the $\Gamma$ of the emission region can be directly estimated by opacity method as
\begin{equation}
\begin{aligned}
    \Gamma_0\simeq\Gamma_{\rm opacity}=&100\left[  \frac{396.9}{C_2(1+z)^{\alpha}} 
    \left(\frac{L_{511\rm keV}}{10^{52}\rm\,erg\,s^{-1}}\right)\right.\\
    &\left(\frac{5.11\rm\,GeV}{E_c}\right) ^{1-\alpha} 
    \left(\frac{-\alpha}{2}\right) ^{-5/3}\\
    &\left.\left(\frac{33.4\rm\,ms}{\delta t}\right) \right] ^{1/(2-2\alpha)},
\end{aligned}
\end{equation}
where $\delta t$ is the minimum variability time, $\alpha$ is the photon index of the high energy component, $L_{511\rm keV}$ is the luminosity at 511\,keV obtained from high energy component, and $C_2\sim1$ \citep{14C_Ajello_2020,C2_Vianello_2018}. 
Considering the temporal profile of the soft non-thermal component and the hard non-thermal components are different, and the lightcurve of \textit{Fermi}/GBM BGO detector in 1-40\,MeV are mainly contributed by the hard component, MVT of this lightcurve is used to calculate $\Gamma_{\rm opacity}$, which is obtained by the baysian block method \citep{typeIL_wang_2025} with a value of 58\,ms. Hence we got $\Gamma_{\rm opacity}\sim$123 for GRB 240825A. On the other hand, the result from opacity should be consistent with self-annihilation that $\Gamma_{\rm opacity}\leq \Gamma_{\rm opacity,max} = (1+z)E_c/m_ec^2$, thus $\Gamma_{\rm opacity,max}$ is $282$ with a $E_c$ of 86.9\,MeV and the estimation $\Gamma_0\sim$123 is adoptable. 

The result from the photosphere method ($\Gamma_{\rm ph}\sim365$) is larger than the result from opacity method ($\Gamma_0\sim123$) is consistent with the picture of fire model in which the kinetic energy is dissipated during the coast phase. 

Moreover, $\Gamma_0\sim123$ is consistent with the constraint of $\Gamma_0\gtrsim52$ for the wind case while conflicting with the constraint of $\Gamma_0\gtrsim139$ for ISM case, which are obtained from optical afterglow \citep{25A_Cheng_2024}, suggests that the wind circumstance, the typical circumstance for stellar collapse, is more favored.

\subsection{Short Duration} 
The $T_{90}$, one of the most classic and effective criterion to classify GRBs, is determined by calculating the time interval between the epochs when the total accumulated net photon counts reach the 5\% and 95\% levels. GRBs can be phenomenologically classified into long GRBs and short GRBs based on the bimodal distribution of $T_{90}$ with a boundary at 2\,s. It should be noted that different instruments and the choice of energy bands can lead to different $T_{90}$.

Despite the classification of GRBs into long and short, a physical classification is based on the progenitor, that the GRBs produced from stellar collapse are named as type-II while the GRBs generated by merger are named as type-I. Long GRBs usually belong to type-II and short GRBs generally belong to type-I. Recently, some short duration GRBs have been shown to be associated with supernovae, thus belongs to genuine type-II GRBs, including GRB 090426 \citep{26_Antonelli_2009}, GRB 200826A \citep{26A_Ahumada_2021,26A_Zhang_2021}, GRB 230812B \citep{12B_Wang_2024}.

\begin{figure*}
\centering
\begin{tabular}{cc}
\begin{minipage}[b]{0.45\linewidth}
    \begin{overpic}[width=\textwidth]{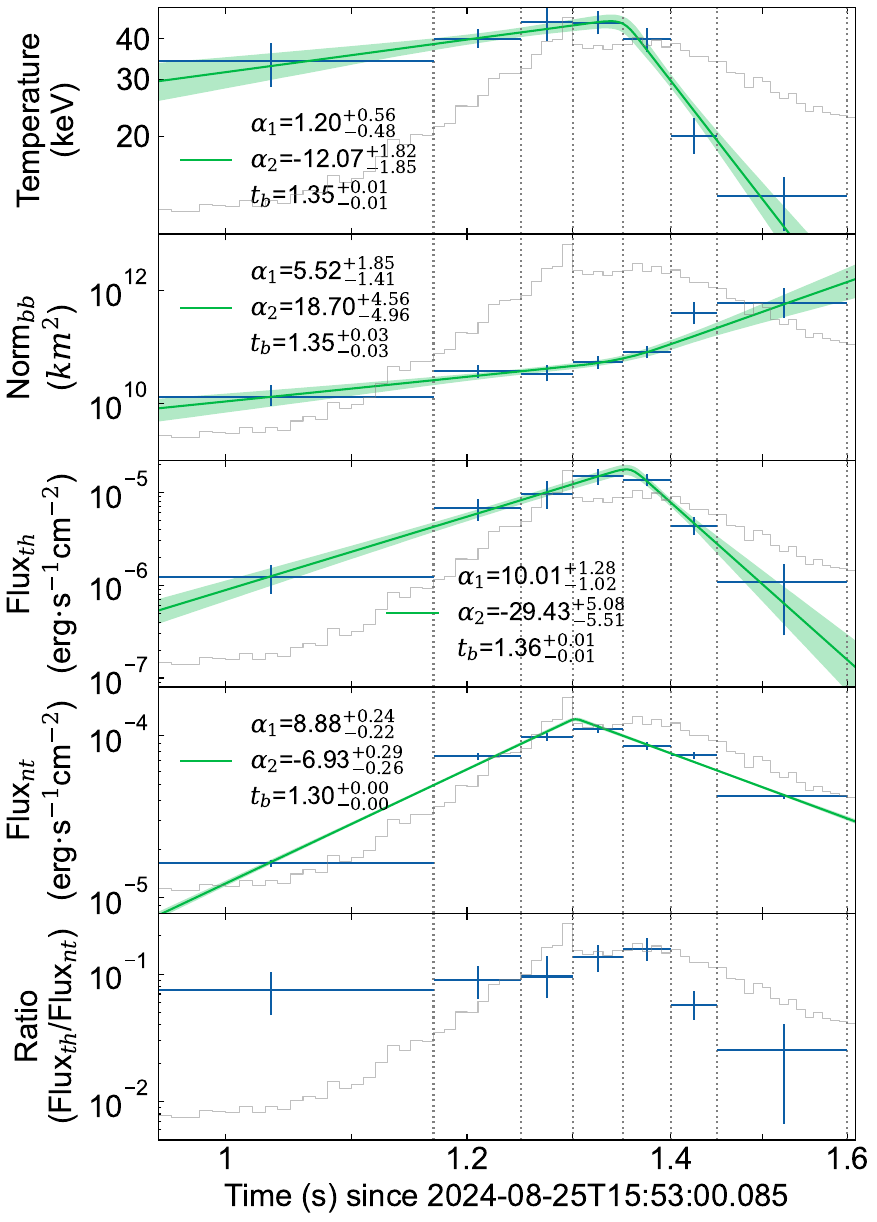}\put(0, 100){\bf a}\end{overpic}
\end{minipage}
\begin{minipage}[b]{0.4\linewidth}
        \begin{overpic}[width=\textwidth]{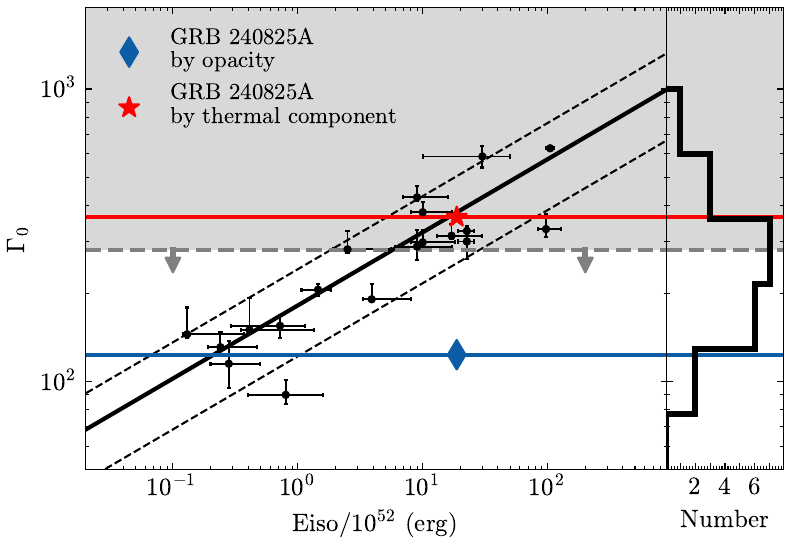}\put(0, 70){\bf b}\end{overpic}
        \begin{overpic}[width=\textwidth]{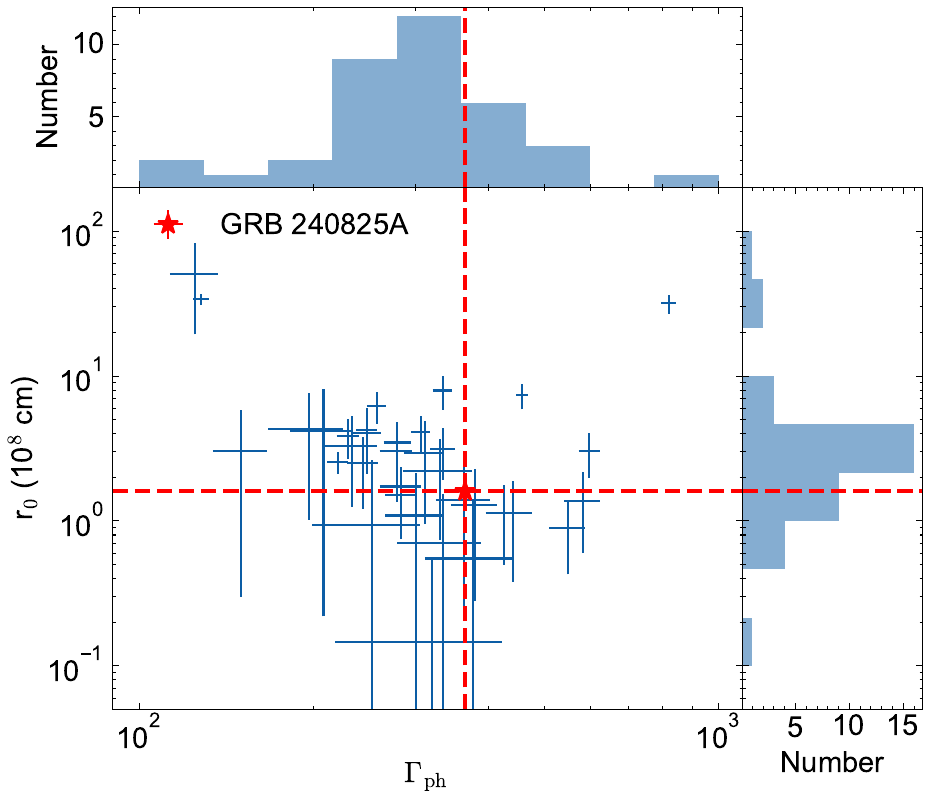}\put(0, 70){\bf c}\end{overpic}
\end{minipage} \\
\begin{overpic}[width=0.4\linewidth]{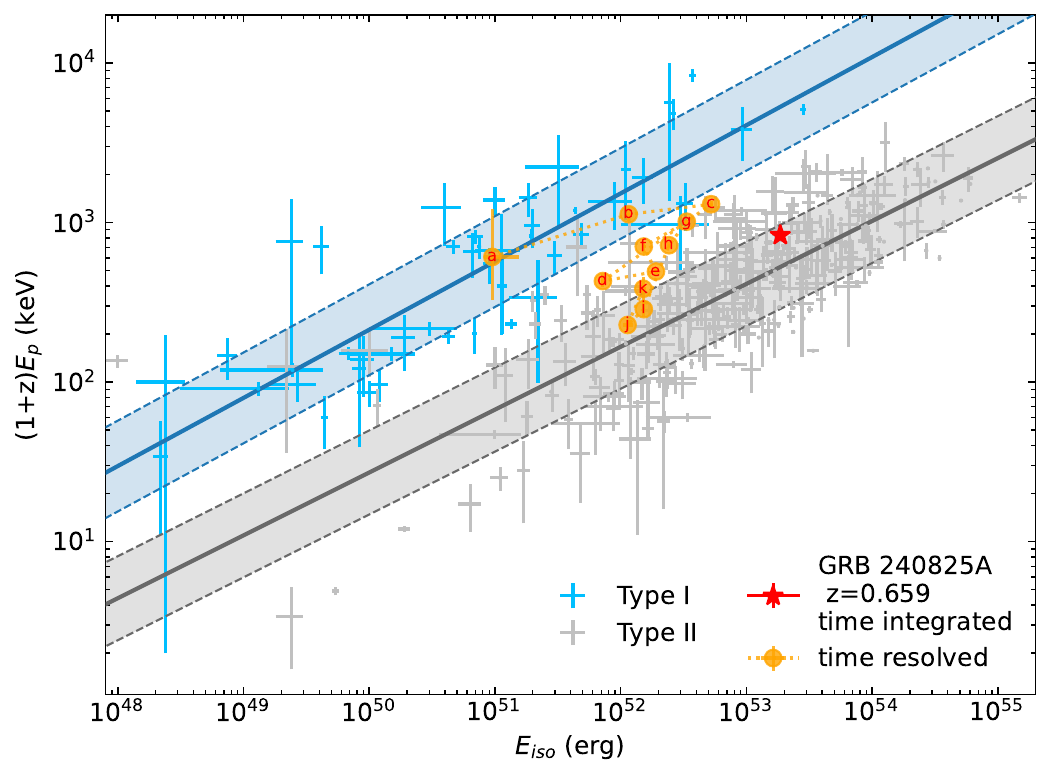}\put(0, 70){\bf d}\end{overpic}
\begin{overpic}[width=0.4\linewidth]{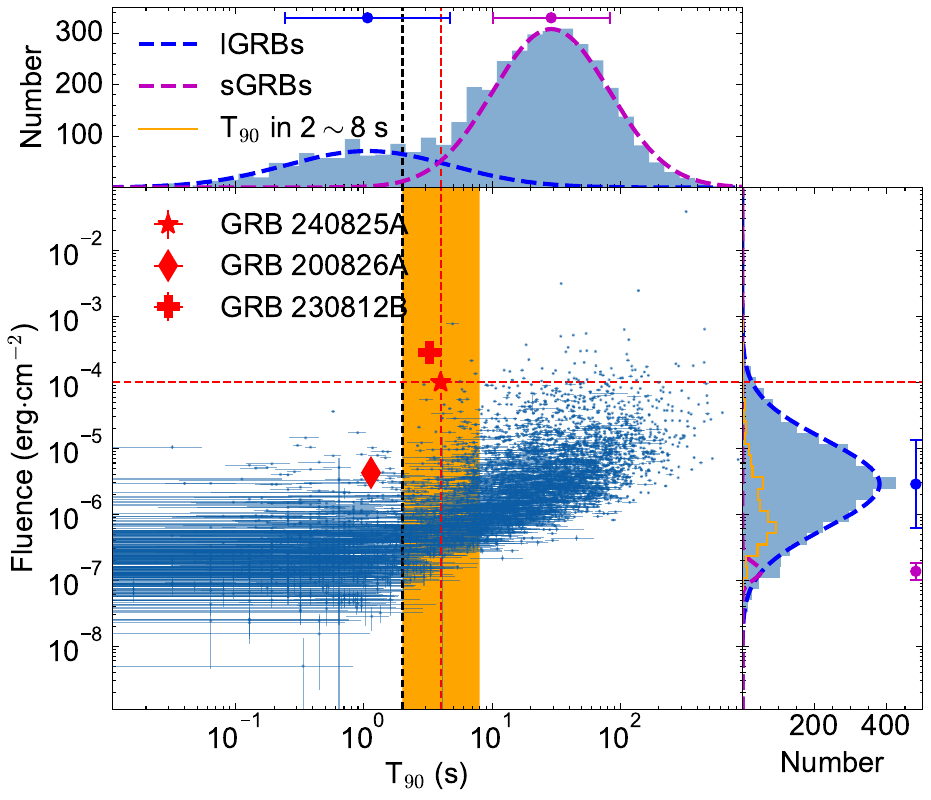}\put(0, 70){\bf e}\end{overpic}\\

\end{tabular}
\caption{\textbf{The evolution of thermal component and the comparison of GRB 240825A with other GRBs.} (a), the broken powerlaw evolution of the thermal component. The flux of thermal component and soft non-thermal component are calculated in energy range of 10 to 1000 keV. (b), $\Gamma$ obtained by different method and location in the $\Gamma_0-\rm E_{iso}$ diagram. The gray dashed line represent the upper limit of $\Gamma_0$ by opacity method. (c) $\Gamma_{\rm ph}$ of GRB 240825A obtained by photosphere emission compared with other samples obtained by the same method. (d), the Amati relation of GRB 240825A. (e), the position of GRB 240825A in the $T_{90}$-Fluence diagram. 
All error bars on data points represent their 1$\sigma$ confidence level. All shaded areas around the best-fit lines represent their 1$\sigma$ confidence bands.}
\label{fig:Gamma}
\end{figure*}

Although there are no reported associated supernovae, the analysis of the host galaxy \citep{25A_Cheng_2024}, the bulk Lorentz factor (see Section \ref{sec:gamma}) and the Amati relation Figure\,\ref{fig:Gamma}(d) suggest the collapsar original of GRB 240825A. 

For the collapsar GRBs, the duration is related to the free-fall timescale of the progenitor, which is excepted to be 
\begin{equation}
    t_{\rm ff}\approx \left(\frac{3\pi}{32G\rho} \right)^{1/2}\approx210\,{\rm s}\left(\frac{\rho}{\rm100\,g\,cm^{-3}} \right)^{-1/2},
    \label{equ:tff}
\end{equation}
with a typical density of massive star as $\rho\approx{\rm100\,g\,cm^{-3}}$, and much longer the duration of GRB 240825A, which can be seen in Figure\,\ref{fig:lc}a that $T_{90}$ of GRB 240825A is calculated for each mission, 

It is also interesting that GRB 240825A exhibits a high brightness that does not match its duration. On one hand, as shown in Figure\,\ref{fig:Gamma}e, the fluence of GRB 240825A exceeds that of 95\% of GRBs detected by \textit{Fermi}/GBM, specifically, GRB 240825A is one of the five brightest GRBs with $T_{90}$ of 2 to 8 (approximation of $T_{90}$ of GRB 240825A obtained by Swift/BAT) currently detected by \textit{Fermi}/GBM. 
On the other hand, the $T_{90}$ of GRB 240825A obtained by \textit{Fermi}/GBM falls within the 1$\sigma$ range of short GRB samples when fitting the $T_{90}$ histogram of the GRB catalog of \textit{Fermi}/GBM by double Gaussian distributions. And $T_{90}$ observed by HXMT and GECAM-C is even $\lesssim$2\,s. 

Therefore, GRB 240825A can be regarded as a short-duration type-II GRB, and the detailed analysis with its high brightness can provide some insights for understanding this type of short-duration ``long" GRBs. Usually two possible scenarios can lead to the short duration of a type-II GRB: a baryon-loaded jet and the jet penetration in the envelope of stellar \citep{26A_Zhang_2021}.

For the former case, the baryon-rich jet is difficult to be accelerated, and will stall in a mildly relativistic or non-relativistic state.
As the $\Gamma$ of GRB 240825A is restricted to a normal level of GRBs (Figure\,\ref{fig:Gamma}b and c), the baryon-loaded jet scenario is disfavored for GRB 240825A.

For the latter case, some observation suggests that the relativistic jet needs some time to go through the stellar envelope (denoted as $\Delta t_{\rm jet}$) before breaking out of the surface and producing the GRB (with a duration of $\Delta t_{\rm GRB}$) \citep{jet_Bromberg_2012}. $\Delta t_{\rm GRB}$ will be short if $\Delta t_{\rm jet}$ is comparable to the central engine activity timescale $\Delta t_{\rm eng}$ as $\Delta t_{\rm GRB}=\Delta t_{\rm eng}-\Delta t_{\rm jet}$. \cite{jet_Bromberg_2011} mentioned that, with a typical values for LGRBs, the time that jet breaks out of the stellar envelope is 
\begin{equation}
\begin{aligned}
    \Delta t_b\approx&15\,{\rm s}\left(\frac{L_{\rm iso}}{10^{51}\,{\rm erg\,s ^{-1}}}\right)^{-\frac{1}{3}}
    \left(\frac{\theta_j}{10^\circ}\right)^{\frac{2}{3}}
    \left(\frac{R_*}{5\,R_\odot}\right)^{\frac{2}{3}}\\
    &\left(\frac{M_*}{15\,M_\odot}\right)^{\frac{1}{3}},
\end{aligned}
\end{equation}
where $L_{\rm iso}$ is the isotropic equivalent jet luminosity and $\Delta\theta_j$ is the jet half-opening angle, $R_*$ and $M_*$ are the radius and the mass of the progenitor star. 
For GRB 240825A, $L_{\rm iso}=1.87^{+0.01}_{-0.01}\times10^{52}\,\rm erg\,s^{-1}$, and $\theta_j\gtrsim0.9^\circ$ (see Section\,\ref{sec:HLE}), that a $\Delta t_b\gtrsim1.13\,\rm s$ is expected and $\Delta t_{\rm eng}\gtrsim6.02\,\rm s$. Considering that $\Delta t_b$ is not very large, with such a high $L_{\rm iso}$, $\Delta t_b$ will not be larger than this result by an order of magnitude. 
When propagating in the stellar envelope, the jet will deposit energy into the stellar envelopes and heat it up, which will lead to a continuous increase in soft X-ray and UV radiation. Moreover, a soft, quasi-thermal X-ray emission will also arise from shock breakout. Therefore, phase \textit{a} of S-II, a faint emission, with a duration that also fits the requirement of $\Delta t_b$, may potentially originate from this stage, but the lack of soft X-ray observation restricts further validation and analysis of this.


\subsection{High latitude emission} \label{sec:HLE}

The lightcurve of GRB X-ray afterglow follows a multi-segment broken powerlaw, including a steep decay phase (tail emission), a shallow decay phase (or plateau), steeper decay phase (pre and post jet break phase), and X-ray flares \citep{ag_zhang_2006}. The earliest segment is a steep decay phase, also known as the tail emission, which is a temporal bridge between the GRB prompt emission and the afterglow emission. The standard convention of the tail emission can be written as 
\begin{equation}
F_\nu\propto t^{-\alpha_t}\nu^{-\beta_t},
\end{equation}
where $\alpha_t$ is the temporal slope and $\beta_t$ is the spectral index \footnote{Here we add a subscript in the temporal decay index and spectral index to distinguish from the photon index $\alpha$ and $\beta$ in BAND model.}. 
The deep slope of the tail emission is dominated by the high latitude emission (HLE), which is also called curvature effect. The HLE refers to the phenomenon of photons emitted from higher latitudes (respect to the line of sight) arriving later than those from lower latitude, causing a sustained but rapidly decaying emission. 
A well-known conclusion resulting from the curvature effect is $\alpha_t=2+\beta_t$ with the assumption of a power law spectrum with spectral index $\beta_t$. Although this relation is no longer proper when the intrinsically curved spectral shape and strong spectral evolution are taken into account, the time-dependent $\alpha_t$ and $\beta_t$ still follow the relation when considering a narrow energy band and the intrinsically curved spectrum can be approximated by a power law \citep{HLE_zhang_2009,07A_sun_2024}.

\begin{figure*}
\centering
\begin{tabular}{c}
\begin{overpic}[width=0.9\textwidth]{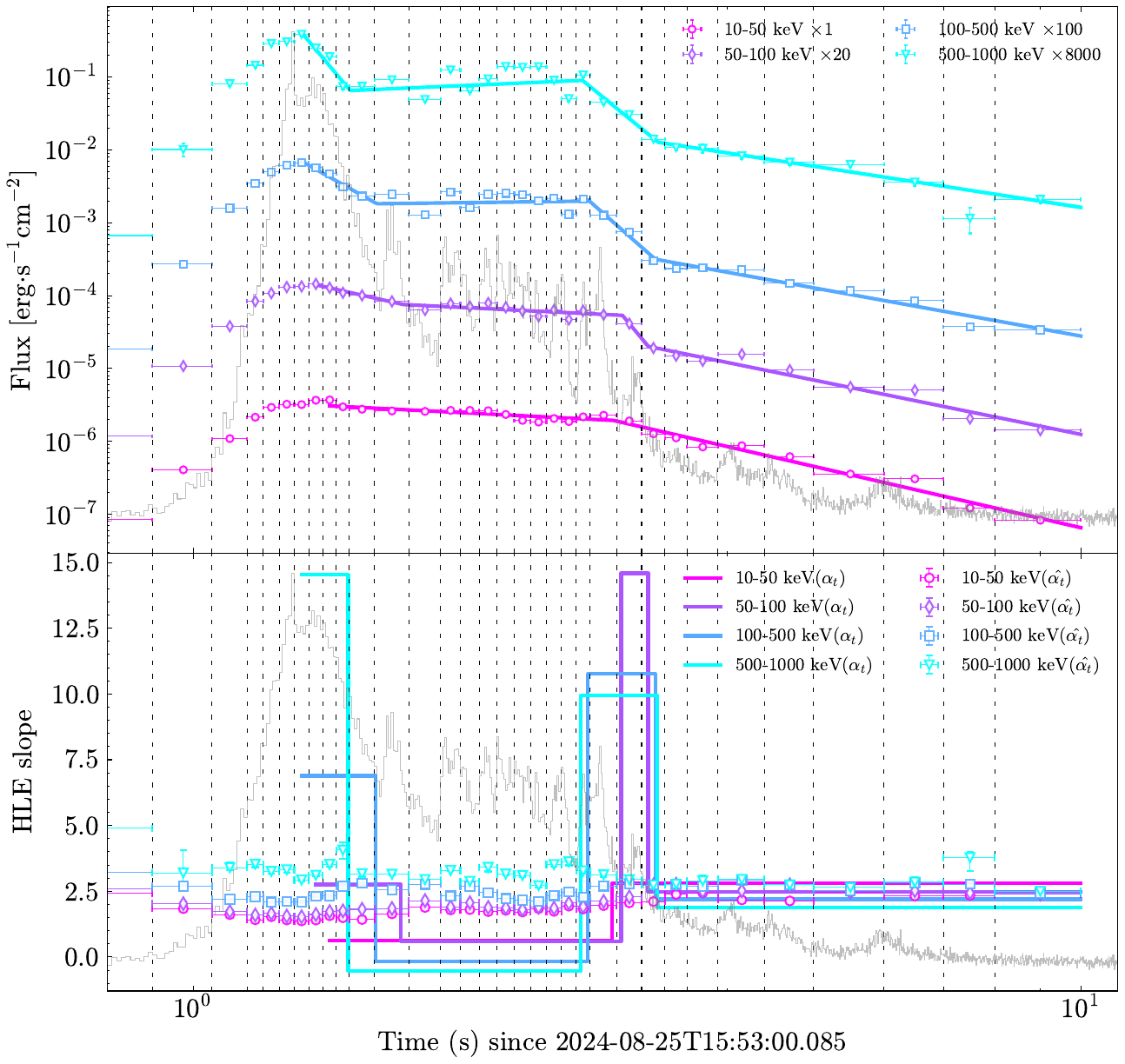}\end{overpic} 
\end{tabular}
\caption{\textbf{The flux lightcurves GRB 240825A.} The top panel is the fitting result of the flux light (based on the spectral fitting result of S-IV spectra) by multi-segment broken powerlaw. The bottom panel is the observed temporal ($\alpha_t$) slopes, which are represent in solid lines, compared with the temporal slopes predicted by the curvature effect ($\hat{\alpha_t}$, shown by data point) in different energy range. The last segment of broken powerlaw evolution of the flux lightcurve in different energy bands roughly conforms to the predictions of high-latitude effects.
All error bars on data points represent their 1$\sigma$ confidence level.}
\label{fig:HLE}
\end{figure*}

The lightcurve of GRB 240825A is a combination of a bright main pulse and a ``plateau-like" structure superimposed on the smooth decay edge of the main pulse. To investigate the nature of the ``plateau-like" structure and the possible tail emission (Figure\,\ref{fig:Gamma}c shows a powerlaw decay of the soft non-thermal component), we conducted the spectral fitting with a finer time slice in 8-8000\,keV (denoted as S-IV, only \textit{Fermi}/GBM, Swift/BAT and GECAM-B data are utilized), and got the high time resolution flux lightcurve, which is shown in the Figure\,\ref{fig:HLE}, then we used multi-segments broken powerlaw to fit the flux lightcurves. A single broken powerlaw (two segments) is expressed as
\begin{equation}
F = \left\{
    \begin{array}{l}
    At^{-\alpha_1}, t<t_{{\rm b}1}\\
    At_{{\rm b}1}^{\alpha_2-\alpha_1}t^{-\alpha_2}, t\geq t_{{\rm b}1}
    \end{array}
    \right.
\end{equation}
where $t_{{\rm b}1}$ is the break time, $\alpha_1$ and $\alpha_2$ are the slope before and after the break time.
Correspondingly, the four segment broken powerlaw is expressed as
\begin{equation}
F = \left\{
    \begin{array}{l}
    At^{-\alpha_1}, t<t_{{\rm b}1}\\
    At_{{\rm b}1}^{\alpha_2-\alpha_1}t^{-\alpha_2}, t_{{\rm b}1}\leq t<t_{{\rm b}2}\\
    At_{{\rm b}1}^{\alpha_2-\alpha_1}t_{{\rm b}2}^{\alpha_3-\alpha_2}t^{-\alpha_3}, t_{{\rm b}2}\leq t<t_{{\rm b}3}\\
    At_{{\rm b}1}^{\alpha_2-\alpha_1}t_{{\rm b}2}^{\alpha_3-\alpha_2}t_{{\rm b}3}^{\alpha_4-\alpha_3}t^{-\alpha_4}, t\geq t_{{\rm b}3}\\
    \end{array}
    \right.
\end{equation}
The flux lightcurves of 50-100\,keV, 100-500\,keV and 500-1000\,keV are well fitted by four segments broken powerlaw from the peak time to the end, well the flux lightcurve of 10-50\,keV are less structured, and can be adequately fitted by broken powerlaw. The fitting results are list in Table\,\ref{tab:lc_fit}. The evolution of the temporal slope is shown in the bottom panel of Figure\,\ref{fig:HLE} as the solid lines, and $\alpha_t$ is the slope of each segment of the broken powerlaw. Following \cite{07A_sun_2024}, $\beta_t$ is calculated by the average spectral index $-\Delta log_{10}F_\nu/\Delta log_{10}E$, and the predicted temporal slope by HLE (denoted as $\hat{\alpha_t}$) are calculated by $\hat{\alpha_t}=\beta_t+2$, which is displayed as the data point in the bottom panel of Figure\,\ref{fig:HLE}. 

The results show that during the last segment of broken powerlaw of the flux lightcurve in all energy bands, the slope is roughly consistent with the prediction of curvature effect, which suggests an early afterglow has already started since $\sim$3.5\,s. 
The slope of the 10-50 keV flux lightcurve at all times does not exceed the upper limit of the slope of the tail emission given by the curvature effect. 
In comparison, the slopes of the flux lightcurves in the other three energy bands (50-100\,keV, 100-500\,keV and 500-1000\,keV) during the first and third segments of powerlaw evolution significantly exceed the predictions of curvature effects, but it returns to the predictions of curvature effects eventually. 

Usually the curvature effects give the upper limit for the temporal slope during the tail emission, but if the jet is very narrow, a possible scenario is that the photons from the outermost layer of the shell with a radius $R_{\rm GRB}$ has reached the observer while the flux is still not become the dominated component, then a sharp decay will be seen \citep{07A_sun_2024,ag_zhang_2006}. If the duration of tail emission is obtained, the opening angle can be derived as 
\begin{equation}
\theta_j=\sqrt{\frac{2c\Delta t_b}{(1+z)R_{\rm GRB}}}   
\label{equ:angle}
\end{equation}
with the assumption of a radius $R_{\rm GRB}$. 

In such a scenario, the behavior of the multi-segment of broken powerlaw flux lightcurve suggests a possible structured jet with two cores for GRB 240825A. 
One narrow jet, which produced a narrow pulse and ended tail emission very early (before $t_{{\rm b}1}$), while the other wider jet only started to enter tail emission at a later time (after $t_{{\rm b}3}$). With Equation\,\ref{equ:angle}, an upper limit of the narrow jet opening angle is $\theta_{j,\rm n}\lesssim0.4^\circ(R_{\rm GRB}/10^{15}\rm cm)^{-\frac{1}{2}}$, by assuming a typical radius of $R_{\rm GRB}$ = $10^{15}\rm cm$. Correspondingly, a lower limit of the wide jet opening angle is $\theta_{j,\rm n}\gtrsim0.9^\circ(R_{\rm GRB}/10^{15}\rm cm)^{-\frac{1}{2}}$.

\section{Summary and Conclusion} \label{sec:conl}
In this paper, we conducted a comprehensive spectral and temporal analysis of GRB 240825A by utilizing the observations from \textit{Fermi}/GBM, GECAM, Swift/BAT, \textit{Fermi}/LAT, and \textit{Inisght}-HXMT. 

Through spectral fitting with different time resolutions, we identify three spectral components in this GRB: a dominant soft non-thermal component well described by the BAND model, a hard non-thermal component well described by the CPL model, and a subdominant thermal component.

The evolution of these three components in GRB 240825A exhibits several intriguing and unusual characteristics:

\begin{enumerate}
    \item The subdominated thermal component appears in the early stage of the prompt emission but disappears before the beginning of the GeV emission.
    \item The GeV emission is bright but has a relatively low cutoff energy.
    \item The BAND component spans a broad energy range and even extends beyond the $E_{\rm c}$ of the hard non-thermal component.
\end{enumerate}

By analyzing the evolution of these three spectral components, we obtained the bulk Lorentz factor at the photosphere ($\Gamma_{\rm ph}\sim365$) and the bulk Lorentz factor before deceleration using the opacity method ($\Gamma_0\sim123$). These values align with the GRB fireball model, where a fraction of the kinetic energy is dissipated during the prompt emission phase. This result provides another valuable sample for investigating GRB ejecta composition and fireball dynamics. Additionally, the evolution of the soft and hard components indicates that the observed positive-to-negative transition of spectral lag is actually an incorrect pulse alignment of CCF.   

We also noticed that the duration of GRB 240825A lies near the boundary between long and short GRBs and is relatively short, given its brightness. 
The high Lorentz factor ($\Gamma$) disfavors the baryon-loaded jet scenario, instead favoring the jet penetration scenario.

Finally, we found that the decay phase of GRB 240825A is well described by a multi-segment powerlaw, including a plateau structure commonly associated with magnetar emergence in other GRBs. 
The temporal and spectral characteristics of the final segment are consistent with high-latitude emission. 
Based on the duration of high-latitude emission, we estimate that the opening angle of the GRB jet must be at least greater than 0.9$^\circ$. Meanwhile, the first segment, exhibiting a very steep slope in the energy range above 100 keV, suggests the presence of a second, narrower jet with an opening angle smaller than 0.4$^\circ$.

\section*{Acknowledgments}
We appreciate Zhuo Li, Qiang Wang, Rui-Ze Shi for the helpful discussion and suggestions that improved this work. 
This work is supported by the National Key R\&D Program of China (2021YFA0718500), 
the Strategic Priority Research Program, the Chinese Academy of Sciences (Grant No. 
XDA30050000, 
XDB0550300
),
and the National Natural Science Foundation of China (Grant No. 12273042, 12494572). 
Rahim Moradi acknowledges support from the Chinese Academy of Sciences (E32984U810). 
The GECAM (Huairou-1) mission is supported by the Strategic Priority Research Program on Space Science (Grant No. XDA15360000) of Chinese Academy of Sciences. 
This work made use of data from the \textit{Insight}-HXMT mission, funded by the CNSA and CAS.
We appreciate the public data and software of \textit{Fermi} and Swift.


\begin{longrotatetable}
\centering
\begin{deluxetable*}{@{}ccccccccccccccccccc@{}}
\tablecaption{Time resolved spectra fitting result of GRB 240825A with different time resolution}\label{tab:spec_fit}
\tabletypesize{\scriptsize}
\renewcommand{\arraystretch}{0.2}
\setlength{\tabcolsep}{3pt}
\tablehead{
 & & & & \multicolumn{4}{c}{soft NT component} & \multicolumn{4}{c}{hard NT component} & \multicolumn{2}{c}{TH component} &  & \\
 \cline{4-14 }
& start & end & model combination & $\alpha$ & $\beta$ & $E_{\rm p}$ & $\log_{10}A$ & $\alpha$ & $\beta$ & $E_{\rm c}$ & $\log_{10}A$ & kT & $\log_{10}A$ & stat/dof & BIC & best model \\
& (s) & (s) & & & & (keV) & & & & (MeV) & & (keV) & & & & 
}
\renewcommand{\arraystretch}{0.9}
\startdata
\cline{1-17}
\textbf{S-I} \\
 & \multirow{4}{*}{0} & \multirow{4}{*}{10} &BAND+CPL+BB & -0.90$^{+0.02}_{-0.02}$ & -2.40$^{+0.02}_{-0.02}$ & 502.08$^{+22.87}_{-19.27}$ & -0.84$^{+0.01}_{-0.02}$ & -1.25$^{+0.04}_{-0.04}$ & -- &79.14$^{+7.61}_{-7.45}$ & -0.24$^{+0.17}_{-0.18}$ & 34.91$^{+2.02}_{-2.32}$ & -1.60$^{+0.12}_{-0.08}$ &902/665&961&\multirow{4}{*}{BAND+CPL+BB}\\
\cline{4-16}
& & & BAND+CPL &-0.82$^{+0.01}_{-0.01}$ & -2.36$^{+0.02}_{-0.03}$ & 404.93$^{+9.20}_{-10.12}$ & -0.76$^{+0.01}_{-0.01}$ & -1.26$^{+0.08}_{-0.05}$ & -- &71.79$^{+9.54}_{-8.80}$ & -0.17$^{+0.15}_{-0.32}$ &-- &-- &923/667&969& \\
\cline{4-16}
& & & BAND+BAND+BB &-0.91$^{+0.02}_{-0.02}$ & -2.48$^{+0.07}_{-0.10}$ & 520.46$^{+25.23}_{-24.45}$ & -0.85$^{+0.02}_{-0.02}$ & -1.19$^{+0.08}_{-0.09}$ & -2.48$^{+0.07}_{-0.10}$ & 56.56$^{+21.45}_{-10.76}$ & -2.62$^{+0.15}_{-0.17}$ & 33.84$^{+2.22}_{-1.97}$ & -1.52$^{+0.08}_{-0.08}$ &899/664&964& \\
\cline{4-16}
& & & BAND+BAND &-0.81$^{+0.01}_{-0.01}$ & -2.36$^{+0.03}_{-0.03}$ & 403.86$^{+13.24}_{-9.89}$ & -0.76$^{+0.01}_{-0.01}$ & -1.30$^{+0.08}_{-0.05}$ & -2.36$^{+0.03}_{-0.03}$ & 76.49$^{+13.78}_{-13.22}$ & -2.59$^{+0.12}_{-0.20}$  &-- &-- &923/666&975& \\
\cline{1-17}
\textbf{S-II} \\
 & \multirow{7}{*}{-0.3} & \multirow{7}{*}{0.9}& CPL & -1.49$^{+0.14}_{-0.09}$ & -- &366.81$^{+361.16}_{-170.13}$ & 0.84$^{+0.14}_{-0.18}$ & -- & -- & -- & -- & -- & -- &393/437&411&\multirow{7}{*}{CPL}\\
\cline{4-16}
& & & BAND & \multicolumn{12}{c}{Unconstrained} \\
\cline{4-16}
& & & BAND+CPL & \multicolumn{12}{c}{Unconstrained} \\
\cline{4-16}
& & & CPL+CPL & \multicolumn{12}{c}{Unconstrained} \\
\cline{4-16}
& & & BAND+BAND & \multicolumn{12}{c}{Unconstrained} \\
\cline{4-16}
& & & BB+BAND & \multicolumn{12}{c}{Unconstrained} \\
\cline{4-16}
& & & BB+CPL & \multicolumn{12}{c}{Unconstrained} \\
\cline{2-17}
 & \multirow{7}{*}{0.9} & \multirow{7}{*}{1.2}& CPL & -0.58$^{+0.03}_{-0.03}$ & -- &562.07$^{+27.27}_{-25.07}$ & 0.64$^{+0.06}_{-0.06}$ & -- & -- & -- & -- & -- & -- &391/435&421&\multirow{7}{*}{BAND+BB}\\
\cline{4-16}
& & & BAND & -0.54$^{+0.04}_{-0.04}$ & -2.88$^{+0.20}_{-0.28}$ & 514.34$^{+31.90}_{-29.95}$ & -0.50$^{+0.02}_{-0.02}$ & -- & -- & -- & -- & -- & -- &378/434&415\\
\cline{4-16}
& & & BAND+CPL &-0.28$^{+0.14}_{-0.10}$ & -2.95$^{+0.41}_{-0.73}$ & 232.04$^{+71.21}_{-35.79}$ & -0.47$^{+0.07}_{-0.05}$ & -0.46$^{+0.05}_{-0.08}$ & -- &0.58$^{+0.12}_{-0.05}$ & -0.04$^{+0.10}_{-0.13}$ & -- & -- &364/431&419\\
\cline{4-16}
& & & CPL+CPL &-0.29$^{+0.11}_{-0.11}$ & -- &221.01$^{+60.04}_{-34.59}$ & 0.07$^{+0.17}_{-0.18}$ &-0.45$^{+0.08}_{-0.09}$ & -- &0.56$^{+0.09}_{-0.07}$ & -0.02$^{+0.18}_{-0.17}$ & -- & -- &366/432&414\\
\cline{4-16}
& & & BAND+BAND & \multicolumn{12}{c}{Unconstrained} \\
\cline{4-16}
& & & BB+BAND &-0.64$^{+0.06}_{-0.06}$ & -3.16$^{+0.27}_{-0.34}$ & 679.33$^{+75.68}_{-58.88}$ & -0.64$^{+0.04}_{-0.04}$ & -- & -- & -- & -- &33.96$^{+4.33}_{-3.81}$ & -0.97$^{+0.13}_{-0.13}$&365/432&413\\
\cline{4-16}
& & & BB+CPL &-0.68$^{+0.06}_{-0.06}$ & -- &740.58$^{+75.00}_{-66.02}$ & 0.70$^{+0.10}_{-0.10}$ & -- & -- & -- & -- &37.78$^{+4.39}_{-4.74}$ & -1.04$^{+0.16}_{-0.16}$&369/433&412\\
\cline{2-17}
 & \multirow{7}{*}{1.2} & \multirow{7}{*}{1.5}& CPL & -0.59$^{+0.01}_{-0.01}$ & -- &681.21$^{+14.38}_{-14.55}$ & 1.21$^{+0.03}_{-0.03}$ & -- & -- & -- & -- & -- & -- &1016/435&1047&\multirow{7}{*}{BAND+BB}\\
\cline{4-16}
& & & BAND & -0.43$^{+0.02}_{-0.02}$ & -2.51$^{+0.04}_{-0.04}$ & 504.48$^{+14.48}_{-14.65}$ & 0.10$^{+0.01}_{-0.01}$ & -- & -- & -- & -- & -- & -- &684/434&720\\
\cline{4-16}
& & & BAND+CPL &-0.11$^{+0.05}_{-0.05}$ & -2.45$^{+0.06}_{-0.08}$ & 259.67$^{+21.66}_{-13.26}$ & 0.19$^{+0.02}_{-0.03}$ & -0.48$^{+0.04}_{-0.04}$ & -- &0.80$^{+0.08}_{-0.06}$ & 0.37$^{+0.08}_{-0.09}$ & -- & -- &579/431&634\\
\cline{4-16}
& & & CPL+CPL &-0.21$^{+0.06}_{-0.05}$ & -- &317.23$^{+20.17}_{-19.71}$ & 0.54$^{+0.09}_{-0.12}$ &-0.60$^{+0.06}_{-0.07}$ & -- &1.19$^{+0.09}_{-0.10}$ & 0.59$^{+0.18}_{-0.22}$ & -- & -- &607/432&655\\
\cline{4-16}
& & & BAND+BAND & \multicolumn{12}{c}{Unconstrained} \\
\cline{4-16}
& & & BB+BAND &-0.59$^{+0.03}_{-0.03}$ & -2.78$^{+0.07}_{-0.08}$ & 784.83$^{+33.05}_{-37.74}$ & -0.11$^{+0.02}_{-0.02}$ & -- & -- & -- & -- &36.74$^{+1.79}_{-1.89}$ & -0.29$^{+0.07}_{-0.07}$&560/432&609\\
\cline{4-16}
& & & BB+CPL &-0.72$^{+0.02}_{-0.02}$ & -- &1073.33$^{+40.57}_{-36.81}$ & 1.25$^{+0.04}_{-0.04}$ & -- & -- & -- & -- &41.90$^{+1.50}_{-1.53}$ & -0.36$^{+0.05}_{-0.05}$&660/433&702\\
\cline{2-17}
 & \multirow{7}{*}{1.5} & \multirow{7}{*}{1.6}& CPL & \multicolumn{12}{c}{Unconstrained} &\multirow{6}{*}{BAND+BB}\\
\cline{4-16}
& & & BAND & \multicolumn{12}{c}{Unconstrained} \\
\cline{4-16}
& & & BAND+CPL &-0.18$^{+0.06}_{-0.06}$ & -2.46$^{+0.12}_{-0.15}$ & 204.36$^{+13.28}_{-10.01}$ & 0.22$^{+0.03}_{-0.04}$ & -0.84$^{+0.09}_{-0.07}$ & -- &2.09$^{+0.74}_{-0.70}$ & 0.32$^{+0.17}_{-0.25}$ & -- & -- &478/431&533\\
\cline{4-16}
& & & CPL+CPL &-0.19$^{+0.04}_{-0.04}$ & -- &210.95$^{+11.43}_{-10.06}$ & 0.54$^{+0.06}_{-0.06}$ &-0.83$^{+0.05}_{-0.05}$ & -- &1.78$^{+0.35}_{-0.29}$ & 0.66$^{+0.10}_{-0.12}$ & -- & -- &486/432&534\\
\cline{4-16}
& & & BAND+BAND & \multicolumn{12}{c}{Unconstrained} \\
\cline{4-16}
& & & BB+BAND &-0.20$^{+0.07}_{-0.09}$ & -2.30$^{+0.06}_{-0.07}$ & 258.89$^{+27.25}_{-20.81}$ & 0.10$^{+0.05}_{-0.07}$ & -- & -- & -- & -- &16.46$^{+3.03}_{-2.65}$ & 0.35$^{+0.19}_{-0.24}$&482/432&530\\
\cline{4-16}
& & & BB+CPL &-1.02$^{+0.07}_{-0.05}$ & -- &1047.23$^{+212.26}_{-215.14}$ & 1.56$^{+0.08}_{-0.10}$ & -- & -- & -- & -- &35.20$^{+1.61}_{-1.59}$ & -0.14$^{+0.06}_{-0.06}$&521/433&563\\
\cline{2-17}
 & \multirow{7}{*}{1.6} & \multirow{7}{*}{1.9}& CPL & \multicolumn{12}{c}{Unconstrained} &\multirow{6}{*}{BAND+CPL}\\
\cline{4-16}
& & & BAND & \multicolumn{12}{c}{Unconstrained} \\
\cline{4-16}
& & & BAND+CPL &-0.65$^{+0.04}_{-0.03}$ & -2.19$^{+0.05}_{-0.07}$ & 297.19$^{+16.90}_{-20.30}$ & -0.13$^{+0.03}_{-0.02}$ & -1.12$^{+0.10}_{-0.08}$ & -- &65.28$^{+20.43}_{-11.93}$ & 0.21$^{+0.23}_{-0.39}$ & -- & -- &518/431&572\\
\cline{4-16}
& & & CPL+CPL &-0.71$^{+0.04}_{-0.03}$ & -- &373.79$^{+13.30}_{-13.52}$ & 1.19$^{+0.06}_{-0.07}$ &-1.26$^{+0.05}_{-0.04}$ & -- &63.63$^{+8.13}_{-6.61}$ & 1.18$^{+0.13}_{-0.20}$ & -- & -- &546/432&594\\
\cline{4-16}
& & & BAND+BAND &-0.69$^{+0.03}_{-0.03}$ & -2.23$^{+0.06}_{-0.09}$ & 318.05$^{+19.28}_{-15.66}$ & -0.15$^{+0.02}_{-0.02}$ & -1.04$^{+0.18}_{-0.12}$ & -4.49$^{+0.88}_{-1.27}$ & 56.55$^{+11.70}_{-11.86}$ & -2.17$^{+0.23}_{-0.35}$ & -- & -- &518/430&579\\
\cline{4-16}
& & & BB+BAND & \multicolumn{12}{c}{Unconstrained} \\
\cline{4-16}
& & & BB+CPL & \multicolumn{12}{c}{Unconstrained} \\
\cline{2-17}\\
\\
 & \multirow{7}{*}{1.9} & \multirow{7}{*}{2.1}& CPL & \multicolumn{12}{c}{Unconstrained} &\multirow{6}{*}{CPL+CPL}\\
\cline{4-16}
& & & BAND & \multicolumn{12}{c}{Unconstrained} \\
\cline{4-16}
& & & BAND+CPL &-0.73$^{+0.04}_{-0.04}$ & -4.65$^{+1.44}_{-1.93}$ & 427.86$^{+18.00}_{-19.82}$ & -0.23$^{+0.02}_{-0.02}$ & -1.23$^{+0.07}_{-0.04}$ & -- &45.42$^{+8.95}_{-6.98}$ & 1.15$^{+0.15}_{-0.25}$ & -- & -- &473/431&528\\
\cline{4-16}
& & & CPL+CPL &-0.71$^{+0.03}_{-0.03}$ & -- &426.37$^{+20.54}_{-19.44}$ & 1.19$^{+0.05}_{-0.05}$ &-1.27$^{+0.03}_{-0.04}$ & -- &47.91$^{+8.29}_{-5.84}$ & 1.31$^{+0.11}_{-0.12}$ & -- & -- &474/432&522\\
\cline{4-16}
& & & BAND+BAND & \multicolumn{12}{c}{Unconstrained} \\
\cline{4-16}
& & & BB+BAND & \multicolumn{12}{c}{Unconstrained} \\
\cline{4-16}
& & & BB+CPL & \multicolumn{12}{c}{Unconstrained} \\
\cline{2-17}
 & \multirow{7}{*}{2.1} & \multirow{7}{*}{2.5}& CPL & \multicolumn{12}{c}{Unconstrained} &\multirow{6}{*}{BAND+CPL}\\
\cline{4-16}
& & & BAND & \multicolumn{12}{c}{Unconstrained} \\
\cline{4-16}
& & & BAND+CPL &-0.71$^{+0.02}_{-0.02}$ & -3.26$^{+0.33}_{-0.59}$ & 610.52$^{+18.44}_{-17.51}$ & -0.27$^{+0.01}_{-0.01}$ & -1.28$^{+0.07}_{-0.05}$ & -- &59.98$^{+9.16}_{-8.25}$ & 0.90$^{+0.18}_{-0.28}$ & -- & -- &489/431&544\\
\cline{4-16}
& & & CPL+CPL &-0.67$^{+0.04}_{-0.04}$ & -- &606.45$^{+21.41}_{-20.25}$ & 1.06$^{+0.08}_{-0.08}$ &-1.32$^{+0.07}_{-0.03}$ & -- &54.36$^{+8.27}_{-7.49}$ & 1.25$^{+0.14}_{-0.28}$ & -- & -- &495/432&544\\
\cline{4-16}
& & & BAND+BAND & \multicolumn{12}{c}{Unconstrained} \\
\cline{4-16}
& & & BB+BAND & \multicolumn{12}{c}{Unconstrained} \\
\cline{4-16}
& & & BB+CPL & \multicolumn{12}{c}{Unconstrained} \\
\cline{2-17}
 & \multirow{7}{*}{2.5} & \multirow{7}{*}{3.0}& CPL & \multicolumn{12}{c}{Unconstrained} &\multirow{6}{*}{BAND+CPL}\\
\cline{4-16}
& & & BAND & \multicolumn{12}{c}{Unconstrained} \\
\cline{4-16}
& & & BAND+CPL &-0.81$^{+0.02}_{-0.02}$ & -3.01$^{+0.32}_{-0.38}$ & 434.40$^{+19.60}_{-17.88}$ & -0.32$^{+0.01}_{-0.01}$ & -1.24$^{+0.05}_{-0.04}$ & -- &69.95$^{+15.59}_{-11.66}$ & 0.58$^{+0.16}_{-0.22}$ & -- & -- &558/431&613\\
\cline{4-16}
& & & CPL+CPL &-0.81$^{+0.02}_{-0.02}$ & -- &446.27$^{+16.78}_{-16.28}$ & 1.29$^{+0.04}_{-0.04}$ &-1.26$^{+0.06}_{-0.04}$ & -- &61.14$^{+10.71}_{-8.50}$ & 0.80$^{+0.14}_{-0.22}$ & -- & -- &563/432&611\\
\cline{4-16}
& & & BAND+BAND & \multicolumn{12}{c}{Unconstrained} \\
\cline{4-16}
& & & BB+BAND & \multicolumn{12}{c}{Unconstrained} \\
\cline{4-16}
& & & BB+CPL & \multicolumn{12}{c}{Unconstrained} \\
\cline{2-17}
 & \multirow{7}{*}{3.0} & \multirow{7}{*}{3.7}& CPL & \multicolumn{12}{c}{Unconstrained} &\multirow{6}{*}{BAND+CPL}\\
\cline{4-16}
& & & BAND & \multicolumn{12}{c}{Unconstrained} \\
\cline{4-16}
& & & BAND+CPL &-1.06$^{+0.07}_{-0.08}$ & -2.16$^{+0.08}_{-0.16}$ & 172.64$^{+21.49}_{-19.57}$ & -0.75$^{+0.04}_{-0.03}$ & -1.50$^{+0.04}_{-0.03}$ & -- &69.97$^{+9.87}_{-8.92}$ & 1.62$^{+0.11}_{-0.17}$ & -- & -- &437/431&491\\
\cline{4-16}
& & & CPL+CPL &-1.31$^{+0.03}_{-0.02}$ & -- &372.13$^{+52.90}_{-42.83}$ & 1.73$^{+0.04}_{-0.05}$ &-1.44$^{+0.04}_{-0.03}$ & -- &73.58$^{+11.64}_{-10.08}$ & 1.36$^{+0.10}_{-0.12}$ & -- & -- &444/432&493\\
\cline{4-16}
& & & BAND+BAND & \multicolumn{12}{c}{Unconstrained} \\
\cline{4-16}
& & & BB+BAND & \multicolumn{12}{c}{Unconstrained} \\
\cline{4-16}
& & & BB+CPL & \multicolumn{12}{c}{Unconstrained} \\
\cline{2-17}
 & \multirow{7}{*}{3.7} & \multirow{7}{*}{5.0}& CPL & -1.35$^{+0.02}_{-0.02}$ & -- &364.68$^{+41.15}_{-35.12}$ & 1.59$^{+0.04}_{-0.04}$ & -- & -- & -- & -- & -- & -- &512/435&543&\multirow{7}{*}{BAND+CPL}\\
\cline{4-16}
& & & BAND & -1.21$^{+0.05}_{-0.05}$ & -2.11$^{+0.04}_{-0.05}$ & 202.80$^{+29.26}_{-23.45}$ & -0.99$^{+0.04}_{-0.04}$ & -- & -- & -- & -- & -- & -- &482/434&519\\
\cline{4-16}
& & & BAND+CPL &-1.03$^{+0.10}_{-0.07}$ & -2.33$^{+0.14}_{-0.16}$ & 137.47$^{+10.36}_{-10.63}$ & -0.98$^{+0.04}_{-0.06}$ & -1.46$^{+0.09}_{-0.08}$ & -- &29.45$^{+9.45}_{-9.00}$ & 1.13$^{+0.30}_{-0.31}$ & -- & -- &460/431&515\\
\cline{4-16}
& & & CPL+CPL &-0.93$^{+0.06}_{-0.06}$ & -- &85.11$^{+8.25}_{-6.82}$ & 0.96$^{+0.07}_{-0.09}$ &-1.19$^{+0.05}_{-0.03}$ & -- &1.29$^{+0.54}_{-0.21}$ & 0.92$^{+0.07}_{-0.11}$ & -- & -- &470/432&518\\
\cline{4-16}
& & & BAND+BAND & \multicolumn{12}{c}{Unconstrained} \\
\cline{4-16}
& & & BB+BAND & \multicolumn{12}{c}{Unconstrained} \\
\cline{4-16}
& & & BB+CPL & \multicolumn{12}{c}{Unconstrained} \\
\cline{2-17}
 & \multirow{7}{*}{5.0} & \multirow{7}{*}{10.0}& CPL & -1.36$^{+0.04}_{-0.04}$ & -- &628.75$^{+219.17}_{-129.92}$ & 1.02$^{+0.06}_{-0.06}$ & -- & -- & -- & -- & -- & -- &520/435&550&\multirow{7}{*}{CPL+CPL}\\
\cline{4-16}
& & & BAND & -1.33$^{+0.04}_{-0.04}$ & -2.23$^{+0.06}_{-0.09}$ & 482.86$^{+120.06}_{-82.14}$ & -1.68$^{+0.03}_{-0.03}$ & -- & -- & -- & -- & -- & -- &502/434&538\\
\cline{4-16}
& & & BAND+CPL &-1.27$^{+0.03}_{-0.03}$ & -2.89$^{+0.45}_{-0.31}$ & 289.12$^{+43.60}_{-34.22}$ & -1.69$^{+0.03}_{-0.03}$ & -1.35$^{+0.11}_{-0.06}$ & -- &42.44$^{+9.00}_{-5.74}$ & 0.24$^{+0.19}_{-0.42}$ & -- & -- &482/431&536\\
\cline{4-16}
& & & CPL+CPL &-1.00$^{+0.06}_{-0.05}$ & -- &231.91$^{+41.24}_{-37.85}$ & 0.21$^{+0.08}_{-0.11}$ &-1.53$^{+0.04}_{-0.03}$ & -- &51.84$^{+11.80}_{-8.00}$ & 0.98$^{+0.06}_{-0.08}$ & -- & -- &482/432&531\\
\cline{4-16}
& & & BAND+BAND & \multicolumn{12}{c}{Unconstrained} \\
\cline{4-16}
& & & BB+BAND & \multicolumn{12}{c}{Unconstrained} \\
\cline{4-16}
& & & BB+CPL & \multicolumn{12}{c}{Unconstrained} \\
\cline{1-17}
\textbf{S-III} \\
 & \multirow{4}{*}{0.9} & \multirow{4}{*}{1.17}& BB+BAND &-0.60$^{+0.05}_{-0.05}$ & -2.98$^{+0.34}_{-1.12}$ & 527.70$^{+45.57}_{-48.51}$ & -0.64$^{+0.03}_{-0.02}$ & -- & -- & -- & -- &1.00$^{+0.00}_{-0.00}$ & -1.96$^{+0.10}_{-0.03}$&404/398&452&\multirow{4}{*}{BB+CPL}\\
\cline{4-16}
& & & BB+CPL &-0.71$^{+0.06}_{-0.06}$ & -- &718.95$^{+84.53}_{-75.78}$ & 0.64$^{+0.09}_{-0.10}$ & -- & -- & -- & -- &34.16$^{+4.73}_{-5.69}$ & -1.08$^{+0.22}_{-0.16}$&401/399&443\\
\cline{4-16}
& & & BAND &-0.58$^{+0.05}_{-0.04}$ & -2.60$^{+0.21}_{-0.35}$ & 493.03$^{+44.12}_{-44.26}$ & -0.63$^{+0.03}_{-0.02}$ & -- & -- & -- & -- &&404/400&440\\
\cline{4-16}
& & & CPL &-0.62$^{+0.04}_{-0.04}$ & -- &553.52$^{+34.39}_{-29.71}$ & 0.58$^{+0.06}_{-0.06}$ & -- & -- & -- & -- &&414/401&444\\
\cline{2-17}
 & \multirow{4}{*}{1.17} & \multirow{4}{*}{1.25}& BB+BAND &-0.51$^{+0.06}_{-0.06}$ & -4.01$^{+0.65}_{-0.63}$ & 700.94$^{+63.45}_{-68.01}$ & -0.16$^{+0.04}_{-0.04}$ & -- & -- & -- & -- &42.37$^{+6.30}_{-4.81}$ & -0.72$^{+0.11}_{-0.18}$&383/398&431&\multirow{4}{*}{BB+CPL}\\
\cline{4-16}
& & & BB+CPL &-0.52$^{+0.04}_{-0.04}$ & -- &727.87$^{+47.71}_{-44.81}$ & 0.85$^{+0.07}_{-0.08}$ & -- & -- & -- & -- &39.97$^{+2.85}_{-2.42}$ & -0.62$^{+0.10}_{-0.13}$&383/399&425\\
\cline{4-16}
& & & BAND &-0.37$^{+0.04}_{-0.04}$ & -2.95$^{+0.19}_{-0.27}$ & 522.79$^{+29.17}_{-28.31}$ & -0.01$^{+0.02}_{-0.02}$ & -- & -- & -- & -- &&394/400&430\\
\cline{4-16}
& & & CPL &-0.45$^{+0.03}_{-0.03}$ & -- &593.36$^{+23.89}_{-21.75}$ & 0.84$^{+0.06}_{-0.06}$ & -- & -- & -- & -- &&408/401&438\\
\cline{2-17}
 & \multirow{4}{*}{1.25} & \multirow{4}{*}{1.3}& BB+BAND &-0.48$^{+0.07}_{-0.06}$ & -2.90$^{+0.21}_{-0.29}$ & 734.44$^{+81.06}_{-71.15}$ & -0.07$^{+0.04}_{-0.05}$ & -- & -- & -- & -- &45.14$^{+5.32}_{-5.78}$ & -0.68$^{+0.16}_{-0.13}$&485/398&533&\multirow{4}{*}{BB+BAND}\\
\cline{4-16}
& & & BB+CPL &-0.63$^{+0.05}_{-0.04}$ & -- &991.09$^{+92.92}_{-80.31}$ & 1.10$^{+0.06}_{-0.08}$ & -- & -- & -- & -- &51.19$^{+2.67}_{-3.73}$ & -0.65$^{+0.09}_{-0.07}$&516/399&558\\
\cline{4-16}
& & & BAND &-0.35$^{+0.04}_{-0.04}$ & -2.62$^{+0.12}_{-0.14}$ & 542.10$^{+35.97}_{-31.05}$ & 0.10$^{+0.02}_{-0.02}$ & -- & -- & -- & -- &&509/400&545\\
\cline{4-16}
& & & CPL &-0.47$^{+0.03}_{-0.03}$ & -- &671.07$^{+31.21}_{-26.80}$ & 0.98$^{+0.06}_{-0.06}$ & -- & -- & -- & -- &&548/401&578\\
\cline{2-17}
 & \multirow{4}{*}{1.3} & \multirow{4}{*}{1.35}& BB+BAND &-0.57$^{+0.05}_{-0.05}$ & -2.54$^{+0.17}_{-0.19}$ & 1275.51$^{+165.97}_{-155.84}$ & -0.17$^{+0.04}_{-0.03}$ & -- & -- & -- & -- &44.86$^{+4.04}_{-3.58}$ & -0.47$^{+0.11}_{-0.11}$&453/398&501&\multirow{4}{*}{BB+BAND}\\
\cline{4-16}
& & & BB+CPL &-0.72$^{+0.04}_{-0.04}$ & -- &1907.56$^{+143.91}_{-133.95}$ & 1.20$^{+0.07}_{-0.07}$ & -- & -- & -- & -- &53.41$^{+4.25}_{-4.25}$ & -0.59$^{+0.10}_{-0.09}$&478/399&520\\
\cline{4-16}
& & & BAND &-0.35$^{+0.05}_{-0.05}$ & -2.10$^{+0.07}_{-0.08}$ & 613.74$^{+54.85}_{-48.07}$ & 0.09$^{+0.02}_{-0.02}$ & -- & -- & -- & -- &&483/400&519\\
\cline{4-16}
& & & CPL &-0.64$^{+0.02}_{-0.02}$ & -- &1244.74$^{+65.61}_{-58.41}$ & 1.26$^{+0.04}_{-0.05}$ & -- & -- & -- & -- &&605/401&635\\
\cline{2-17}
 & \multirow{4}{*}{1.35} & \multirow{4}{*}{1.4}& BB+BAND &-0.66$^{+0.05}_{-0.05}$ & -2.99$^{+0.28}_{-0.46}$ & 951.06$^{+76.96}_{-94.80}$ & -0.16$^{+0.04}_{-0.03}$ & -- & -- & -- & -- &39.90$^{+3.23}_{-3.51}$ & -0.30$^{+0.11}_{-0.11}$&373/398&421&\multirow{4}{*}{BB+BAND}\\
\cline{4-16}
& & & BB+CPL &-0.74$^{+0.04}_{-0.04}$ & -- &1164.66$^{+101.87}_{-88.61}$ & 1.26$^{+0.08}_{-0.08}$ & -- & -- & -- & -- &42.97$^{+3.07}_{-2.97}$ & -0.34$^{+0.08}_{-0.08}$&380/399&422\\
\cline{4-16}
& & & BAND &-0.36$^{+0.05}_{-0.05}$ & -2.25$^{+0.08}_{-0.10}$ & 443.84$^{+37.37}_{-35.12}$ & 0.16$^{+0.03}_{-0.02}$ & -- & -- & -- & -- &&416/400&452\\
\cline{4-16}
& & & CPL &-0.58$^{+0.03}_{-0.03}$ & -- &705.46$^{+37.54}_{-33.69}$ & 1.21$^{+0.05}_{-0.05}$ & -- & -- & -- & -- &&475/401&505\\
\cline{2-17}
 & \multirow{4}{*}{1.4} & \multirow{4}{*}{1.45}& BB+BAND &-0.38$^{+0.13}_{-0.09}$ & -2.72$^{+0.15}_{-0.18}$ & 521.44$^{+48.71}_{-48.11}$ & -0.04$^{+0.04}_{-0.04}$ & -- & -- & -- & -- &20.10$^{+2.63}_{-2.45}$ & 0.39$^{+0.20}_{-0.19}$&402/398&450&\multirow{4}{*}{BB+BAND}\\
\cline{4-16}
& & & BB+CPL &-0.55$^{+0.06}_{-0.06}$ & -- &670.04$^{+48.11}_{-40.57}$ & 0.97$^{+0.12}_{-0.10}$ & -- & -- & -- & -- &25.16$^{+2.86}_{-2.14}$ & 0.14$^{+0.12}_{-0.16}$&412/399&454\\
\cline{4-16}
& & & BAND &-0.51$^{+0.04}_{-0.04}$ & -2.64$^{+0.15}_{-0.18}$ & 454.21$^{+33.31}_{-30.29}$ & 0.07$^{+0.02}_{-0.02}$ & -- & -- & -- & -- &&431/400&467\\
\cline{4-16}
& & & CPL &-0.60$^{+0.03}_{-0.03}$ & -- &557.06$^{+27.73}_{-25.98}$ & 1.23$^{+0.05}_{-0.05}$ & -- & -- & -- & -- &&454/401&484\\
\cline{2-17}
 & \multirow{4}{*}{1.45} & \multirow{4}{*}{1.6}& BB+BAND &-0.08$^{+0.11}_{-0.13}$ & -2.33$^{+0.06}_{-0.05}$ & 242.95$^{+13.93}_{-10.72}$ & 0.16$^{+0.03}_{-0.03}$ & -- & -- & -- & -- &13.11$^{+1.87}_{-2.89}$ & 0.58$^{+0.27}_{-0.27}$&470/398&518&\multirow{4}{*}{BB+BAND}\\
\cline{4-16}
& & & BB+CPL &-0.80$^{+0.07}_{-0.08}$ & -- &599.37$^{+95.78}_{-65.03}$ & 1.25$^{+0.11}_{-0.09}$ & -- & -- & -- & -- &35.45$^{+2.14}_{-1.68}$ & -0.27$^{+0.05}_{-0.06}$&558/399&600\\
\cline{4-16}
& & & BAND &-0.23$^{+0.05}_{-0.05}$ & -2.28$^{+0.05}_{-0.06}$ & 236.99$^{+11.83}_{-11.27}$ & 0.16$^{+0.03}_{-0.03}$ & -- & -- & -- & -- &&497/400&534\\
\cline{4-16}
& & & CPL &-0.48$^{+0.03}_{-0.03}$ & -- &336.01$^{+9.79}_{-9.24}$ & 0.96$^{+0.05}_{-0.04}$ & -- & -- & -- & -- &&610/401&640\\
\cline{1-17}
\textbf{S-IV} \\
 & \multirow{3}{*}{-0.3} & \multirow{3}{*}{0.9}& CPL & -1.52$^{+0.19}_{-0.12}$ & -- &437.27$^{+2230.71}_{-257.54}$ & 0.86$^{+0.19}_{-0.23}$ & -- & -- & -- & -- & -- & -- &387/396&417&\multirow{3}{*}{CPL}\\
\cline{4-16}
& & & BAND & -1.37$^{+0.14}_{-0.18}$ & -2.81$^{+0.95}_{-1.49}$ & 173.98$^{+165.17}_{-47.07}$ & -2.08$^{+0.12}_{-0.09}$ & -- & -- & -- & -- & -- & -- &385/395&421\\
\cline{4-16}
& & & PL & -1.65$^{+0.05}_{-0.05}$ & -- &-- &1.04$^{+0.10}_{-0.09}$ & -- & -- & -- & -- & -- & -- &393/397&417\\
\cline{2-17}
 & \multirow{3}{*}{0.9} & \multirow{3}{*}{1.05}& CPL & -0.89$^{+0.08}_{-0.08}$ & -- &454.20$^{+117.65}_{-80.39}$ & 0.69$^{+0.13}_{-0.13}$ & -- & -- & -- & -- & -- & -- &472/396&502&\multirow{3}{*}{BAND}\\
\cline{4-16}
& & & BAND & -0.77$^{+0.25}_{-0.13}$ & -2.46$^{+0.61}_{-1.64}$ & 348.22$^{+141.42}_{-145.43}$ & -1.03$^{+0.19}_{-0.08}$ & -- & -- & -- & -- & -- & -- &410/395&446\\
\cline{4-16}
& & & PL & -1.36$^{+0.03}_{-0.03}$ & -- &-- &1.40$^{+0.06}_{-0.06}$ & -- & -- & -- & -- & -- & -- &493/397&517\\
\cline{2-17}
 & \multirow{3}{*}{1.05} & \multirow{3}{*}{1.15}& CPL & -0.53$^{+0.05}_{-0.05}$ & -- &566.23$^{+35.90}_{-34.63}$ & 0.58$^{+0.09}_{-0.08}$ & -- & -- & -- & -- & -- & -- &391/396&421&\multirow{3}{*}{CPL}\\
\cline{4-16}
& & & BAND & -0.49$^{+0.06}_{-0.05}$ & -2.97$^{+0.44}_{-0.87}$ & 516.78$^{+48.10}_{-53.44}$ & -0.45$^{+0.03}_{-0.03}$ & -- & -- & -- & -- & -- & -- &388/395&424\\
\cline{4-16}
& & & PL & -1.26$^{+0.01}_{-0.01}$ & -- &-- &1.79$^{+0.03}_{-0.03}$ & -- & -- & -- & -- & -- & -- &928/397&952\\
\cline{2-17}
 & \multirow{3}{*}{1.15} & \multirow{3}{*}{1.2}& CPL & -0.48$^{+0.05}_{-0.04}$ & -- &556.71$^{+35.11}_{-31.71}$ & 0.82$^{+0.08}_{-0.09}$ & -- & -- & -- & -- & -- & -- &375/396&405&\multirow{3}{*}{BAND}\\
\cline{4-16}
& & & BAND & -0.36$^{+0.06}_{-0.06}$ & -2.60$^{+0.18}_{-0.22}$ & 453.91$^{+36.11}_{-36.41}$ & -0.08$^{+0.03}_{-0.03}$ & -- & -- & -- & -- & -- & -- &357/395&393\\
\cline{4-16}
& & & PL & -1.23$^{+0.01}_{-0.01}$ & -- &-- &2.04$^{+0.02}_{-0.03}$ & -- & -- & -- & -- & -- & -- &1048/397&1072\\
\cline{2-17}
 & \multirow{3}{*}{1.2} & \multirow{3}{*}{1.25}& CPL & -0.47$^{+0.03}_{-0.04}$ & -- &626.48$^{+29.06}_{-26.52}$ & 0.93$^{+0.07}_{-0.06}$ & -- & -- & -- & -- & -- & -- &388/396&418&\multirow{3}{*}{CPL}\\
\cline{4-16}
& & & BAND & -0.43$^{+0.05}_{-0.04}$ & -3.35$^{+0.42}_{-0.82}$ & 579.96$^{+38.66}_{-37.56}$ & -0.00$^{+0.02}_{-0.02}$ & -- & -- & -- & -- & -- & -- &389/395&425\\
\cline{4-16}
& & & PL & -1.21$^{+0.01}_{-0.01}$ & -- &-- &2.13$^{+0.02}_{-0.02}$ & -- & -- & -- & -- & -- & -- &1301/397&1325\\
\cline{2-17}
 & \multirow{3}{*}{1.25} & \multirow{3}{*}{1.3}& CPL & -0.47$^{+0.03}_{-0.03}$ & -- &672.23$^{+30.42}_{-29.25}$ & 0.98$^{+0.06}_{-0.06}$ & -- & -- & -- & -- & -- & -- &544/396&574&\multirow{3}{*}{BAND}\\
\cline{4-16}
& & & BAND & -0.34$^{+0.05}_{-0.04}$ & -2.47$^{+0.11}_{-0.13}$ & 522.71$^{+33.91}_{-32.51}$ & 0.10$^{+0.02}_{-0.02}$ & -- & -- & -- & -- & -- & -- &498/395&534\\
\cline{4-16}
& & & PL & -1.18$^{+0.01}_{-0.01}$ & -- &-- &2.16$^{+0.02}_{-0.02}$ & -- & -- & -- & -- & -- & -- &1659/397&1683\\
\cline{2-17}
 & \multirow{3}{*}{1.3} & \multirow{3}{*}{1.35}& CPL & -0.64$^{+0.02}_{-0.02}$ & -- &1232.48$^{+57.57}_{-58.03}$ & 1.25$^{+0.05}_{-0.05}$ & -- & -- & -- & -- & -- & -- &578/396&607&\multirow{3}{*}{BAND}\\
\cline{4-16}
& & & BAND & -0.30$^{+0.06}_{-0.05}$ & -1.96$^{+0.07}_{-0.08}$ & 550.52$^{+51.66}_{-49.24}$ & 0.11$^{+0.03}_{-0.02}$ & -- & -- & -- & -- & -- & -- &471/395&507\\
\cline{4-16}
& & & PL & -1.13$^{+0.01}_{-0.01}$ & -- &-- &2.08$^{+0.02}_{-0.02}$ & -- & -- & -- & -- & -- & -- &1445/397&1469\\
\cline{2-17}
 & \multirow{3}{*}{1.35} & \multirow{3}{*}{1.4}& CPL & -0.58$^{+0.03}_{-0.03}$ & -- &710.16$^{+36.20}_{-34.21}$ & 1.22$^{+0.05}_{-0.05}$ & -- & -- & -- & -- & -- & -- &474/396&504&\multirow{3}{*}{BAND}\\
\cline{4-16}
& & & BAND & -0.32$^{+0.06}_{-0.06}$ & -2.13$^{+0.09}_{-0.10}$ & 408.90$^{+40.94}_{-37.02}$ & 0.18$^{+0.03}_{-0.03}$ & -- & -- & -- & -- & -- & -- &404/395&440\\
\cline{4-16}
& & & PL & -1.21$^{+0.01}_{-0.01}$ & -- &-- &2.23$^{+0.02}_{-0.02}$ & -- & -- & -- & -- & -- & -- &1479/397&1503\\
\cline{2-17}
 & \multirow{3}{*}{1.4} & \multirow{3}{*}{1.45}& CPL & -0.60$^{+0.03}_{-0.03}$ & -- &554.28$^{+27.55}_{-25.29}$ & 1.23$^{+0.05}_{-0.06}$ & -- & -- & -- & -- & -- & -- &482/396&512&\multirow{3}{*}{BAND}\\
\cline{4-16}
& & & BAND & -0.50$^{+0.05}_{-0.04}$ & -2.55$^{+0.14}_{-0.18}$ & 445.38$^{+33.41}_{-32.86}$ & 0.08$^{+0.02}_{-0.02}$ & -- & -- & -- & -- & -- & -- &462/395&498\\
\cline{4-16}
& & & PL & -1.26$^{+0.01}_{-0.01}$ & -- &-- &2.29$^{+0.02}_{-0.02}$ & -- & -- & -- & -- & -- & -- &1372/397&1396\\
\cline{2-17}
 & \multirow{3}{*}{1.45} & \multirow{3}{*}{1.5}& CPL & -0.40$^{+0.05}_{-0.05}$ & -- &326.09$^{+15.95}_{-14.02}$ & 0.87$^{+0.08}_{-0.07}$ & -- & -- & -- & -- & -- & -- &450/396&480&\multirow{3}{*}{BAND}\\
\cline{4-16}
& & & BAND & -0.33$^{+0.07}_{-0.06}$ & -2.91$^{+0.26}_{-0.33}$ & 293.69$^{+19.68}_{-21.54}$ & 0.11$^{+0.04}_{-0.03}$ & -- & -- & -- & -- & -- & -- &437/395&473\\
\cline{4-16}
& & & PL & -1.33$^{+0.01}_{-0.01}$ & -- &-- &2.32$^{+0.02}_{-0.02}$ & -- & -- & -- & -- & -- & -- &1403/397&1427\\
\cline{2-17}
 & \multirow{3}{*}{1.5} & \multirow{3}{*}{1.6}& CPL & -0.52$^{+0.03}_{-0.03}$ & -- &342.06$^{+14.21}_{-12.77}$ & 1.02$^{+0.06}_{-0.06}$ & -- & -- & -- & -- & -- & -- &613/396&643&\multirow{3}{*}{BAND}\\
\cline{4-16}
& & & BAND & -0.20$^{+0.07}_{-0.07}$ & -2.14$^{+0.05}_{-0.06}$ & 215.39$^{+14.21}_{-13.01}$ & 0.18$^{+0.04}_{-0.04}$ & -- & -- & -- & -- & -- & -- &483/395&519\\
\cline{4-16}
& & & PL & -1.34$^{+0.01}_{-0.01}$ & -- &-- &2.31$^{+0.02}_{-0.02}$ & -- & -- & -- & -- & -- & -- &1739/397&1763\\
\cline{2-17}
 & \multirow{3}{*}{1.6} & \multirow{3}{*}{1.75}& CPL & -0.68$^{+0.03}_{-0.03}$ & -- &475.22$^{+19.81}_{-18.51}$ & 1.20$^{+0.05}_{-0.05}$ & -- & -- & -- & -- & -- & -- &581/396&611&\multirow{3}{*}{BAND}\\
\cline{4-16}
& & & BAND & -0.56$^{+0.06}_{-0.05}$ & -2.19$^{+0.09}_{-0.10}$ & 350.36$^{+30.86}_{-34.66}$ & -0.09$^{+0.03}_{-0.03}$ & -- & -- & -- & -- & -- & -- &492/395&528\\
\cline{4-16}
& & & PL & -1.32$^{+0.01}_{-0.01}$ & -- &-- &2.23$^{+0.02}_{-0.02}$ & -- & -- & -- & -- & -- & -- &1758/397&1782\\
\cline{2-17}
 & \multirow{3}{*}{1.75} & \multirow{3}{*}{1.9}& CPL & -1.01$^{+0.03}_{-0.03}$ & -- &466.80$^{+36.09}_{-33.53}$ & 1.67$^{+0.05}_{-0.05}$ & -- & -- & -- & -- & -- & -- &513/396&543&\multirow{3}{*}{BAND}\\
\cline{4-16}
& & & BAND & -0.77$^{+0.05}_{-0.05}$ & -1.95$^{+0.05}_{-0.05}$ & 233.36$^{+20.94}_{-20.52}$ & -0.20$^{+0.04}_{-0.03}$ & -- & -- & -- & -- & -- & -- &420/395&456\\
\cline{4-16}
& & & PL & -1.44$^{+0.01}_{-0.01}$ & -- &-- &2.33$^{+0.02}_{-0.02}$ & -- & -- & -- & -- & -- & -- &897/397&921\\
\cline{2-17}
 & \multirow{3}{*}{1.9} & \multirow{3}{*}{2.0}& CPL & -0.76$^{+0.03}_{-0.03}$ & -- &593.10$^{+30.89}_{-30.66}$ & 1.30$^{+0.05}_{-0.05}$ & -- & -- & -- & -- & -- & -- &444/396&474&\multirow{3}{*}{BAND}\\
\cline{4-16}
& & & BAND & -0.73$^{+0.03}_{-0.03}$ & -2.61$^{+0.14}_{-0.18}$ & 545.79$^{+33.01}_{-30.80}$ & -0.20$^{+0.01}_{-0.02}$ & -- & -- & -- & -- & -- & -- &423/395&459\\
\cline{4-16}
& & & PL & -1.31$^{+0.01}_{-0.01}$ & -- &-- &2.19$^{+0.02}_{-0.02}$ & -- & -- & -- & -- & -- & -- &1179/397&1203\\
\cline{2-17}
 & \multirow{3}{*}{2.0} & \multirow{3}{*}{2.1}& CPL & -0.90$^{+0.04}_{-0.03}$ & -- &467.47$^{+38.23}_{-33.13}$ & 1.52$^{+0.06}_{-0.06}$ & -- & -- & -- & -- & -- & -- &563/396&593&\multirow{3}{*}{BAND}\\
\cline{4-16}
& & & BAND & -0.71$^{+0.06}_{-0.06}$ & -1.93$^{+0.06}_{-0.07}$ & 268.80$^{+33.95}_{-28.62}$ & -0.16$^{+0.04}_{-0.04}$ & -- & -- & -- & -- & -- & -- &461/395&496\\
\cline{4-16}
& & & PL & -1.39$^{+0.01}_{-0.01}$ & -- &-- &2.29$^{+0.02}_{-0.02}$ & -- & -- & -- & -- & -- & -- &895/397&919\\
\cline{2-17}
 & \multirow{3}{*}{2.1} & \multirow{3}{*}{2.2}& CPL & -0.73$^{+0.03}_{-0.03}$ & -- &500.71$^{+27.38}_{-25.29}$ & 1.27$^{+0.05}_{-0.05}$ & -- & -- & -- & -- & -- & -- &452/396&482&\multirow{3}{*}{BAND}\\
\cline{4-16}
& & & BAND & -0.67$^{+0.04}_{-0.04}$ & -2.42$^{+0.12}_{-0.14}$ & 420.46$^{+30.23}_{-30.45}$ & -0.16$^{+0.02}_{-0.02}$ & -- & -- & -- & -- & -- & -- &419/395&455\\
\cline{4-16}
& & & PL & -1.33$^{+0.01}_{-0.01}$ & -- &-- &2.22$^{+0.02}_{-0.02}$ & -- & -- & -- & -- & -- & -- &1172/397&1196\\
\cline{2-17}
 & \multirow{3}{*}{2.2} & \multirow{3}{*}{2.3}& CPL & -0.74$^{+0.03}_{-0.03}$ & -- &665.60$^{+36.50}_{-33.95}$ & 1.22$^{+0.05}_{-0.05}$ & -- & -- & -- & -- & -- & -- &510/396&540&\multirow{3}{*}{BAND}\\
\cline{4-16}
& & & BAND & -0.74$^{+0.03}_{-0.03}$ & -2.84$^{+0.20}_{-0.28}$ & 648.03$^{+36.54}_{-38.24}$ & -0.27$^{+0.01}_{-0.01}$ & -- & -- & -- & -- & -- & -- &491/395&527\\
\cline{4-16}
& & & PL & -1.28$^{+0.01}_{-0.01}$ & -- &-- &2.10$^{+0.02}_{-0.02}$ & -- & -- & -- & -- & -- & -- &1130/397&1154\\
\cline{2-17}
 & \multirow{3}{*}{2.3} & \multirow{3}{*}{2.4}& CPL & -0.67$^{+0.03}_{-0.03}$ & -- &658.25$^{+34.25}_{-30.92}$ & 1.03$^{+0.05}_{-0.05}$ & -- & -- & -- & -- & -- & -- &476/396&506&\multirow{3}{*}{BAND}\\
\cline{4-16}
& & & BAND & -0.66$^{+0.03}_{-0.03}$ & -3.73$^{+0.49}_{-0.70}$ & 642.79$^{+34.19}_{-31.87}$ & -0.30$^{+0.01}_{-0.01}$ & -- & -- & -- & -- & -- & -- &433/395&469\\
\cline{4-16}
& & & PL & -1.26$^{+0.01}_{-0.01}$ & -- &-- &2.01$^{+0.02}_{-0.02}$ & -- & -- & -- & -- & -- & -- &1291/397&1315\\
\cline{2-17}
 & \multirow{3}{*}{2.4} & \multirow{3}{*}{2.5}& CPL & -0.87$^{+0.03}_{-0.03}$ & -- &1177.41$^{+92.50}_{-85.62}$ & 1.29$^{+0.05}_{-0.05}$ & -- & -- & -- & -- & -- & -- &504/396&534&\multirow{3}{*}{BAND}\\
\cline{4-16}
& & & BAND & -0.82$^{+0.04}_{-0.03}$ & -2.54$^{+0.20}_{-0.31}$ & 963.17$^{+105.72}_{-95.34}$ & -0.42$^{+0.02}_{-0.02}$ & -- & -- & -- & -- & -- & -- &490/395&526\\
\cline{4-16}
& & & PL & -1.25$^{+0.01}_{-0.01}$ & -- &-- &1.94$^{+0.02}_{-0.02}$ & -- & -- & -- & -- & -- & -- &921/397&945\\
\cline{2-17}
 & \multirow{3}{*}{2.5} & \multirow{3}{*}{2.6}& CPL & -0.68$^{+0.04}_{-0.03}$ & -- &515.28$^{+29.93}_{-26.73}$ & 1.09$^{+0.06}_{-0.06}$ & -- & -- & -- & -- & -- & -- &377/396&407&\multirow{3}{*}{BAND}\\
\cline{4-16}
& & & BAND & -0.65$^{+0.04}_{-0.04}$ & -2.81$^{+0.22}_{-0.36}$ & 476.62$^{+30.16}_{-29.96}$ & -0.26$^{+0.02}_{-0.02}$ & -- & -- & -- & -- & -- & -- &365/395&401\\
\cline{4-16}
& & & PL & -1.31$^{+0.01}_{-0.01}$ & -- &-- &2.09$^{+0.02}_{-0.02}$ & -- & -- & -- & -- & -- & -- &1037/397&1061\\
\cline{2-17}
\\
\\
 & \multirow{3}{*}{2.6} & \multirow{3}{*}{2.7}& CPL & -0.88$^{+0.04}_{-0.04}$ & -- &457.71$^{+35.90}_{-30.82}$ & 1.32$^{+0.06}_{-0.06}$ & -- & -- & -- & -- & -- & -- &415/396&445&\multirow{3}{*}{CPL}\\
\cline{4-16}
& & & BAND & -0.86$^{+0.04}_{-0.04}$ & -3.09$^{+0.46}_{-0.97}$ & 432.37$^{+38.94}_{-35.73}$ & -0.42$^{+0.02}_{-0.02}$ & -- & -- & -- & -- & -- & -- &412/395&448\\
\cline{4-16}
& & & PL & -1.40$^{+0.01}_{-0.01}$ & -- &-- &2.15$^{+0.02}_{-0.02}$ & -- & -- & -- & -- & -- & -- &828/397&852\\
\cline{2-17}
 & \multirow{3}{*}{2.7} & \multirow{3}{*}{2.8}& CPL & -0.76$^{+0.03}_{-0.03}$ & -- &587.13$^{+36.20}_{-30.37}$ & 1.22$^{+0.05}_{-0.05}$ & -- & -- & -- & -- & -- & -- &468/396&498&\multirow{3}{*}{CPL}\\
\cline{4-16}
& & & BAND & -0.75$^{+0.04}_{-0.04}$ & -3.36$^{+0.48}_{-0.86}$ & 562.72$^{+37.33}_{-38.38}$ & -0.30$^{+0.02}_{-0.02}$ & -- & -- & -- & -- & -- & -- &477/395&513\\
\cline{4-16}
& & & PL & -1.31$^{+0.01}_{-0.01}$ & -- &-- &2.10$^{+0.02}_{-0.02}$ & -- & -- & -- & -- & -- & -- &1072/397&1096\\
\cline{2-17}
 & \multirow{3}{*}{2.8} & \multirow{3}{*}{3.0}& CPL & -0.98$^{+0.03}_{-0.03}$ & -- &450.36$^{+25.17}_{-23.63}$ & 1.55$^{+0.04}_{-0.04}$ & -- & -- & -- & -- & -- & -- &543/396&573&\multirow{3}{*}{BAND}\\
\cline{4-16}
& & & BAND & -0.89$^{+0.05}_{-0.04}$ & -2.17$^{+0.10}_{-0.14}$ & 331.67$^{+41.49}_{-41.18}$ & -0.34$^{+0.03}_{-0.03}$ & -- & -- & -- & -- & -- & -- &523/395&559\\
\cline{4-16}
& & & PL & -1.43$^{+0.01}_{-0.01}$ & -- &-- &2.28$^{+0.02}_{-0.02}$ & -- & -- & -- & -- & -- & -- &1126/397&1150\\
\cline{2-17}
 & \multirow{3}{*}{3.0} & \multirow{3}{*}{3.2}& CPL & -1.16$^{+0.03}_{-0.03}$ & -- &498.48$^{+56.76}_{-45.75}$ & 1.73$^{+0.05}_{-0.05}$ & -- & -- & -- & -- & -- & -- &443/396&473&\multirow{3}{*}{BAND}\\
\cline{4-16}
& & & BAND & -0.89$^{+0.08}_{-0.08}$ & -1.87$^{+0.05}_{-0.06}$ & 200.06$^{+37.23}_{-27.74}$ & -0.39$^{+0.06}_{-0.06}$ & -- & -- & -- & -- & -- & -- &408/395&444\\
\cline{4-16}
& & & PL & -1.49$^{+0.01}_{-0.01}$ & -- &-- &2.23$^{+0.02}_{-0.02}$ & -- & -- & -- & -- & -- & -- &664/397&688\\
\cline{2-17}
 & \multirow{3}{*}{3.2} & \multirow{3}{*}{3.4}& CPL & -1.51$^{+0.04}_{-0.03}$ & -- &1238.02$^{+730.17}_{-468.34}$ & 2.03$^{+0.06}_{-0.07}$ & -- & -- & -- & -- & -- & -- &453/396&483&\multirow{3}{*}{BAND}\\
\cline{4-16}
& & & BAND & -0.27$^{+0.16}_{-0.14}$ & -1.73$^{+0.03}_{-0.03}$ & 50.22$^{+4.61}_{-3.95}$ & 0.16$^{+0.16}_{-0.13}$ & -- & -- & -- & -- & -- & -- &420/395&456\\
\cline{4-16}
& & & PL & -1.60$^{+0.02}_{-0.02}$ & -- &-- &2.18$^{+0.04}_{-0.04}$ & -- & -- & -- & -- & -- & -- &472/397&496\\
\cline{2-17}
 & \multirow{3}{*}{3.4} & \multirow{3}{*}{3.6}& CPL & -1.62$^{+0.03}_{-0.03}$ & -- &2728.54$^{+2837.07}_{-1321.71}$ & 2.14$^{+0.05}_{-0.05}$ & -- & -- & -- & -- & -- & -- &449/396&479&\multirow{3}{*}{BAND}\\
\cline{4-16}
& & & BAND & -0.34$^{+0.39}_{-0.18}$ & -1.74$^{+0.03}_{-0.03}$ & 38.93$^{+5.07}_{-4.29}$ & 0.14$^{+0.42}_{-0.18}$ & -- & -- & -- & -- & -- & -- &435/395&471\\
\cline{4-16}
& & & PL & -1.66$^{+0.02}_{-0.02}$ & -- &-- &2.20$^{+0.04}_{-0.04}$ & -- & -- & -- & -- & -- & -- &455/397&479\\
\cline{2-17}
 & \multirow{3}{*}{3.6} & \multirow{3}{*}{3.9}& CPL & -1.47$^{+0.05}_{-0.05}$ & -- &861.42$^{+659.18}_{-301.68}$ & 1.80$^{+0.07}_{-0.08}$ & -- & -- & -- & -- & -- & -- &403/396&433&\multirow{3}{*}{BAND}\\
\cline{4-16}
& & & BAND & -1.44$^{+0.05}_{-0.05}$ & -2.27$^{+0.36}_{-1.34}$ & 643.53$^{+473.57}_{-241.76}$ & -1.13$^{+0.04}_{-0.03}$ & -- & -- & -- & -- & -- & -- &398/395&434\\
\cline{4-16}
& & & PL & -1.59$^{+0.02}_{-0.02}$ & -- &-- &1.98$^{+0.04}_{-0.04}$ & -- & -- & -- & -- & -- & -- &422/397&446\\
\cline{2-17}
 & \multirow{3}{*}{3.9} & \multirow{3}{*}{4.4}& CPL & -1.26$^{+0.04}_{-0.04}$ & -- &284.23$^{+40.32}_{-30.89}$ & 1.54$^{+0.06}_{-0.07}$ & -- & -- & -- & -- & -- & -- &444/396&474&\multirow{3}{*}{BAND}\\
\cline{4-16}
& & & BAND & -0.98$^{+0.12}_{-0.09}$ & -1.94$^{+0.06}_{-0.07}$ & 133.66$^{+23.73}_{-21.38}$ & -0.76$^{+0.10}_{-0.08}$ & -- & -- & -- & -- & -- & -- &410/395&446\\
\cline{4-16}
& & & PL & -1.59$^{+0.01}_{-0.01}$ & -- &-- &2.02$^{+0.03}_{-0.03}$ & -- & -- & -- & -- & -- & -- &563/397&587\\
\cline{2-17}
 & \multirow{3}{*}{4.4} & \multirow{3}{*}{5.0}& CPL & -1.44$^{+0.04}_{-0.04}$ & -- &572.23$^{+161.98}_{-108.82}$ & 1.59$^{+0.07}_{-0.06}$ & -- & -- & -- & -- & -- & -- &426/396&456&\multirow{3}{*}{BAND}\\
\cline{4-16}
& & & BAND & -0.50$^{+0.22}_{-0.28}$ & -1.75$^{+0.04}_{-0.05}$ & 59.65$^{+16.28}_{-8.31}$ & -0.41$^{+0.24}_{-0.28}$ & -- & -- & -- & -- & -- & -- &413/395&449\\
\cline{4-16}
& & & PL & -1.60$^{+0.02}_{-0.02}$ & -- &-- &1.85$^{+0.04}_{-0.04}$ & -- & -- & -- & -- & -- & -- &473/397&497\\
\cline{2-17}
 & \multirow{3}{*}{5.0} & \multirow{3}{*}{6.0}& CPL & -1.45$^{+0.03}_{-0.03}$ & -- &1644.59$^{+1043.50}_{-572.12}$ & 1.38$^{+0.05}_{-0.06}$ & -- & -- & -- & -- & -- & -- &470/396&500&\multirow{3}{*}{CPL}\\
\cline{4-16}
& & & BAND & -0.97$^{+0.31}_{-0.14}$ & -1.67$^{+0.04}_{-0.05}$ & 105.76$^{+43.52}_{-33.05}$ & -1.13$^{+0.29}_{-0.13}$ & -- & -- & -- & -- & -- & -- &475/395&511\\
\cline{4-16}
& & & PL & -1.53$^{+0.02}_{-0.02}$ & -- &-- &1.52$^{+0.03}_{-0.03}$ & -- & -- & -- & -- & -- & -- &481/397&505\\
\cline{2-17}
 & \multirow{3}{*}{6.0} & \multirow{3}{*}{7.0}& CPL & -1.41$^{+0.05}_{-0.05}$ & -- &603.83$^{+424.68}_{-184.56}$ & 1.29$^{+0.07}_{-0.07}$ & -- & -- & -- & -- & -- & -- &414/396&444&\multirow{3}{*}{BAND}\\
\cline{4-16}
& & & BAND & -1.29$^{+0.11}_{-0.08}$ & -1.85$^{+0.09}_{-0.12}$ & 250.84$^{+104.24}_{-68.51}$ & -1.45$^{+0.08}_{-0.06}$ & -- & -- & -- & -- & -- & -- &403/395&439\\
\cline{4-16}
& & & PL & -1.57$^{+0.02}_{-0.02}$ & -- &-- &1.52$^{+0.04}_{-0.04}$ & -- & -- & -- & -- & -- & -- &431/397&455\\
\cline{2-17}
 & \multirow{3}{*}{7.0} & \multirow{3}{*}{8.0}& CPL & -1.30$^{+0.10}_{-0.08}$ & -- &356.38$^{+200.71}_{-95.74}$ & 0.73$^{+0.12}_{-0.14}$ & -- & -- & -- & -- & -- & -- &507/396&537&\multirow{3}{*}{CPL}\\
\cline{4-16}
& & & BAND & -1.21$^{+0.04}_{-0.08}$ & -3.25$^{+1.01}_{-1.24}$ & 273.16$^{+85.86}_{-50.81}$ & -1.81$^{+0.05}_{-0.06}$ & -- & -- & -- & -- & -- & -- &506/395&542\\
\cline{4-16}
& & & PL & -1.55$^{+0.03}_{-0.03}$ & -- &-- &1.09$^{+0.06}_{-0.06}$ & -- & -- & -- & -- & -- & -- &528/397&552\\
\cline{2-17}
 & \multirow{3}{*}{8.0} & \multirow{3}{*}{10.0}& CPL & -1.36$^{+0.05}_{-0.04}$ & -- &3002.45$^{+2517.50}_{-1416.21}$ & 0.63$^{+0.09}_{-0.09}$ & -- & -- & -- & -- & -- & -- &422/396&452&\multirow{3}{*}{CPL}\\
\cline{4-16}
& & & BAND & -1.33$^{+0.06}_{-0.05}$ & -3.21$^{+1.17}_{-1.24}$ & 1663.88$^{+885.70}_{-798.50}$ & -2.07$^{+0.05}_{-0.04}$ & -- & -- & -- & -- & -- & -- &421/395&457\\
\cline{4-16}
& & & PL & -1.45$^{+0.03}_{-0.03}$ & -- &-- &0.77$^{+0.07}_{-0.07}$ & -- & -- & -- & -- & -- & -- &431/397&455\\
\cline{2-17}
\enddata
\end{deluxetable*}
\end{longrotatetable}

\begin{table*}[htbp]
\begin{center}
\caption{Flux lightcurve fitting result of GRB 240825A.}\label{tab:lc_fit}%
\begin{tabular*}{\hsize}{@{}@{\extracolsep{\fill}}cccccccccc@{}}
\toprule
Energy range & $\alpha_1$ & $\alpha_2$ & $\alpha_3$ & $\alpha_4$ & $t_{{\rm b}1}$ & $t_{{\rm b}2}$ & $t_{{\rm b}3}$ & $\log_{10}A$ & $\chi^2$/dof\\ 
(keV) &  &  &  &  & (s) & (s) & (s) & (erg$\cdot$cm$^{-2}\cdot$s$^{-1}$) & \\ 
\midrule
10-50& 0.62$^{+0.03}_{-0.03}$ & 2.80$^{+0.03}_{-0.03}$ & -- & -- & 2.97$^{+0.02}_{-0.02}$ & -- & -- & -5.42$^{+0.01}_{-0.01}$ & 247/21\\
50-100& 2.75$^{+0.12}_{-0.12}$ & 0.60$^{+0.03}_{-0.03}$ & 14.59$^{+3.28}_{-2.20}$ & 2.48$^{+0.03}_{-0.03}$ & 1.72$^{+0.02}_{-0.02}$ & 3.04$^{+0.01}_{-0.01}$ & 3.26$^{+0.04}_{-0.03}$ & -4.77$^{+0.02}_{-0.02}$ & 370/18\\
100-500& 6.89$^{+0.19}_{-0.18}$ & -0.16$^{+0.05}_{-0.04}$ & 10.76$^{+0.30}_{-0.31}$ & 2.20$^{+0.04}_{-0.05}$ & 1.61$^{+0.01}_{-0.01}$ & 2.79$^{+0.01}_{-0.01}$ & 3.32$^{+0.01}_{-0.01}$ & -3.31$^{+0.03}_{-0.03}$ & 777/19\\
500-1000& 14.53$^{+0.57}_{-0.57}$ & -0.53$^{+0.07}_{-0.07}$ & 9.95$^{+0.35}_{-0.33}$ & 1.88$^{+0.08}_{-0.08}$ & 1.50$^{+0.01}_{-0.01}$ & 2.74$^{+0.01}_{-0.01}$ & 3.34$^{+0.03}_{-0.03}$ & -2.52$^{+0.08}_{-0.08}$ & 491/19\\
\botrule
\end{tabular*}
\end{center}
\end{table*}

\clearpage

\bibliography{main}{}
\bibliographystyle{aasjournal}

\end{document}